\documentclass{sig-alternate-05-2015}
\usepackage[
  pass,
]{geometry}

\usepackage[utf8]{inputenc}
\usepackage{microtype}

\usepackage{graphicx}

\usepackage{url}      
\usepackage[bookmarks=false]{hyperref}
\usepackage{cleveref}
\usepackage{amsmath}
\usepackage[caption=false]{subfig}

\usepackage{tabularx}
\usepackage{booktabs}

  \newcolumntype{R}[1]{>{\raggedleft\let\newline\\\arraybackslash\hspace{0pt}}m{#1}}
\usepackage{threeparttable}
  
\usepackage{siunitx}
\usepackage{footnote}

\usepackage{comment}
\usepackage{xcolor}

\usepackage{flushend}

\usepackage{enumitem}
\toappear{
   
   {\confname{\the\conf}}   
   \the\confinfo\par        
}
\begin{document}
%
\CopyrightYear{2016}
\conferenceinfo{MM '16,}{October 15 - 19, 2016, Amsterdam, Netherlands}

\clubpenalty=10000
\widowpenalty = 10000

\title{Are Safer Looking Neighborhoods More Lively? \\ A Multimodal Investigation into Urban Life}
%
%
%
%
%

\numberofauthors{6} 
%
\author{
%
%
\alignauthor
Marco De Nadai\\
       \affaddr{FBK and University of Trento}\\
       \affaddr{Trento, Italy}\\
       \email{denadai@fbk.eu}
\alignauthor Radu L. Vieriu\\
       \affaddr{University of Trento}\\
       \affaddr{Trento, Italy}\\
       \email{radulaurentiu.vieriu@unitn.it}
\alignauthor Gloria Zen\\
       \affaddr{University of Trento}\\
       \affaddr{Trento, Italy}\\
       \email{gloria.zen@unitn.it}
\and    
\alignauthor Stefan  Dragicevic\\
        \affaddr{TIM and University of Trento}\\
       \affaddr{Trento, Italy}\\
       \email{stefan.dragicevic@unitn.it}
\alignauthor Nikhil  Naik\\
       \affaddr{MIT Media Lab}\\
       \affaddr{Cambridge, MA}\\
       \email{naik@mit.edu}
\alignauthor Michele Caraviello\\
       \affaddr{TIM}\\
       \affaddr{Trento, Italy}\\
       \email{michele.caraviello@telecomitalia.it} 
\and
\alignauthor Cesar A. Hidalgo\\
       \affaddr{MIT Media Lab}\\
       \affaddr{Cambridge, MA}\\
       \email{hidalgo@mit.edu}
\alignauthor Nicu Sebe\\
       \affaddr{University of Trento}\\
       \affaddr{Trento, Italy}\\
       \email{sebe@disi.unitn.it}
\alignauthor Bruno Lepri\\
       \affaddr{FBK}\\
       \affaddr{Trento, Italy}\\
       \email{lepri@fbk.eu}  
}
%


\maketitle
\begin{abstract}
Policy makers, urban planners, architects, sociologists, and economists are interested in creating urban areas that are both lively and safe. But are the safety and liveliness of neighborhoods independent characteristics? Or are they just two sides of the same coin? In a world where people avoid unsafe looking places, neighborhoods that look unsafe will be less lively, and will fail to harness the natural surveillance of human activity. But in a world where the preference for safe looking neighborhoods is small, the connection between the perception of safety and liveliness will be either weak or nonexistent. In this paper we explore the connection between the levels of activity and the perception of safety of neighborhoods in two major Italian cities by combining mobile phone data (as a proxy for activity or liveliness) with scores of perceived safety estimated using a Convolutional Neural Network trained on a dataset of Google Street View images scored using a crowdsourced visual perception survey. We find that: (i) safer looking neighborhoods are more active than what is expected from their population density, employee density, and distance to the city centre; and (ii) that the correlation between appearance of safety and activity is positive, strong, and significant, for females and people over 50, but negative for people under 30, suggesting that the behavioral impact of perception depends on the demographic of the population. Finally, we use occlusion techniques to identify the urban features that contribute to the appearance of safety, finding that greenery and street facing windows contribute to a positive appearance of safety (in agreement with Oscar Newman's defensible space theory). These results suggest that urban appearance modulates levels of human activity and, consequently, a neighborhood's rate of natural surveillance. 

\end{abstract}





\section{Introduction}

Does a neighborhood's appearance of safety affect how active it is? For decades scholars from a variety of disciplines, but mainly from urban planning, have been exploring the potential connection between a neighborhood's appearance of safety and its levels of human activity.

The modern literature connecting safety, liveliness, and architecture, can be traced back to Jane Jacobs' seminal 1961 book: \emph{The Death and Life of Great American Cities}~\cite{jacobs1961death}. In there, Jacobs introduced the eyes-on-the-street, or natural surveillance hypothesis~\cite{doeksen1997reducing}, which suggests that citizens can maintain the safety of their neighborhoods naturally through continued surveillance. For natural surveillance to take place, however, Jacobs argued that neighborhoods needed to have certain physical qualities, such as well lit streets and buildings with street facing windows.

Jacobs' idea that the physical quality of a neighborhood can enhance its safety was later expanded by Oscar Newman's \emph{defensible space theory}~\cite{newman1972defensible}. Defensible space theory expands on the idea of natural surveillance by suggesting that neighbors will be more likely to protect an area when there are clear physical demarcations separating what is considered public and private property~\cite{jacobsdefensible, newman1972defensible}. Examples of architectural markers of defensible space are archways in the entrance of building complexes, or staircases in the entrance of townhouses. These archways and staircases do not only serve an aesthetic purpose, but also, signal the boundary between a city's public space and the private and semi-private spaces that neighbors are expected to watch and defend.

Here, we strengthen the link between Jacobs' and Newman's theories by asking whether safer looking neighborhoods are more likely to experience more human activity--and hence, experience more natural surveillance. We explore this connection, by combining computer vision methods, that can be used to measure the physical characteristics of neighborhoods~\cite{naik2015people,quercia2014aesthetic,salesses2013collaborative,porzi2015predicting}, with mobile phone data, which has become a common proxy for human activity~\cite{eagle2010,gonzalez2008,hidalgo2008dynamics,isaacman2010tale,Lenormand150449}, for two Italian cities (Rome and Milan). The combination of computer vision and mobile phone data helps us test whether safer looking neighborhoods are more active, and therefore, if neighborhoods that look physically safer could be experiencing more natural surveillance. 

Our data provides support for a connection between appearance and activity. Using spatially filtered multivariate regressions we find that neighborhoods that are perceived as safer are more active than what is expected from their population density, the density of employees, and their distance to the city center. 
Also, we find that the perception of safety appears to modulate the relative population of females and adults, with unsafer looking neighborhoods experiencing a lower number of female  and people over 50 than safer looking neighborhoods. 
Conversely, we find that younger populations are disproportionately more active in unsafe looking neighborhoods. Finally, we use occlusion techniques to identify the areas of an image that trigger a positive or negative evaluation of safety in the Artificial Neural Network, finding that greenery and street facing windows tend to be associated with higher levels of safety, as suggested by Oscar Newman's defensible space theory.  These observational results strongly suggest--but don't causally prove--that the appearance of neighborhoods has an effect on their levels of human activity, and potentially, on a neighborhood's level of natural surveillance. 

\section{Related Work}

The connection between urban perception and human activity speaks primarily to two streams of literature. The first one is the stream of literature focused on the environmental factors contributing to crime, which has a long tradition in criminology and urban sociology. While our paper does not focus on crime per se, the connection between the physical appearance of neighborhoods and natural surveillance suggested by Jacobs and Newman makes our results relevant to that stream of literature~\cite{jacobs1961death,jacobsdefensible,newman1972defensible}. The second one is the stream of literature using surveys, and more recently, computer vision methods to quantify people’s perception of urban environments~\cite{naik2014streetscore,nasar1998evaluative,salesses2013collaborative}.

\subsection{Neighborhood appearance and crime}

Beyond Jacob's and Newman's theories, the most widely known theory suggesting a connection between urban perception and crime is the \emph{broken windows theory} (BWT) of Wilson and Kelling~\cite{kelling1997fixing,wilson1982broken}. The BWT is the hypothesis that urban incivilities, such as broken windows and litter, promote criminal activity. The classical mechanism used to justify the theory says that urban incivilities signal lawlessness, and may cause the offenders of small incivilities to scale their criminal behavior to more predatory forms of crime if they are not reigned in. The policy implications of the BWT, however, vary from community policing—the promotion of ties between police officers and their communities—to zero-tolerance polices, which promote cracking down on all minor offences to deter more serious forms of crime.

Evidence in favor of the broken windows theory has been presented by Kelling and Coles~\cite{kelling1997fixing}, who looked at data and stories from New York to Seattle to argue that community policing is an effective way to deter more serious forms of crime. Kelling and Sousa~\cite{kelling2001police} provide additional evidence by using an extensive dataset on crime, demographics, and economic data from New York City. More recently, Corman and Mocan~\cite{corman2005carrots} used New York City data on the policing of misdemeanors (as a proxy for broken windows policing), and on robbery, car theft, and grand larceny, to provide evidence in support of broken windows policing. 

The broken windows theory is also supported by a few field experiments, such as those conducted by Keizer \emph{et al.} in The Netherlands~\cite{keizer2008spreading}. In six experiments, Keizer \emph{et al.} intervened environments by spraying graffiti on walls, or leaving supermarket carts unattended and studied the behavior of subjects, in both the presence and absence of disorder, to see when people broke norms (such as littering). Their data showed a significant increase in people’s norm breaking behavior when they were in the presence of disorder. 

But not all of the evidence collected to test the BWT, and its policy implications, is favorable to it~\cite{harcourt2009illusion}. In a 2006 paper, Bernard Harcourt re-analyzed the data presented by Kelling and Sousa~\cite{kelling2001police} and found no evidence of the effectiveness of the broken windows policing~\cite{harcourt2006broken}. More recently, Harcourt and Ludwig used a dataset of more than fifty thousand marijuana related arrests to provide evidence that community policing is not only ineffective, but that it also unfairly targets minorities~\cite{harcourt2007reefer}.

Moreover, the BWT has been criticized by work showing that the social and ethnic context of a neighborhood may matter more than urban disorder. In Sampson \emph{et al.}~\cite{sampson1997neighborhoods} and Sampson and Raudenbush~\cite{sampson2001disorder,sampson2004seeing}, community data from Chicago was used to argue that racial and economic context were more predictive of disorderly behavior than physical disorder. To help bridge their results with the literature, Sampson and Raudenbush~\cite{sampson2001disorder} proposed an alternative interpretation of the theory, where both neighborhood disorder and crime, are manifestations of a lack of informal forms of control within disengaged and distrusting communities. The reframed theory, therefore, interprets the link between disorder and crime as a manifestation of the lack of informal forms of social control and organization.

\subsection{The social and computational image of the city}

The second literature our paper speaks to is the literature measuring urban perception and understand its social and economic implications. The original literature on this topic can be traced back to seminal work by the urbanist Kevin Lynch \cite{lynch1960image}, who interviewed people in Boston, Jersey City, and Los Angeles, to understand the large scale image of cities that people made in their heads. This work on a city's imageability was later continued by social psychologists like Stanley Milgram \cite{milgram1970experience}, and by urbanists like Jack Nasar \cite{nasar1998evaluative, nasar1993proximate}, who created evaluative maps of cities, also using survey methods. 

More recently, however, this literature started leveraging crowdsourcing \cite{salesses2013collaborative} and computer vision methods \cite{naik2014streetscore, porzi2015predicting} to improve the scale, precision, and resolution of the evaluative maps created. 

On the data collection side of this literature, Salesses \emph{et al.}~\cite{salesses2013collaborative} created a large crowdsourced visual perception survey to measure people's perception of streetscapes, and to create comparable evaluative maps for New York, Boston, Linz, and Salzburg, which they also used to measure the segregation and inequality of experiences in these cities, and to show that violent crime correlates with the variance of appearence of safety in an area. 

These new sources of crowd-sourced data gave rise to studies looking to understand the features of an image that explain how streetscapes are perceived. On a recent study, Quercia \emph{et al.}~\cite{quercia2014aesthetic} investigated which visual aspects of London neighborhoods make them appear beautiful, quiet and/or happy. A related study by Porzi \emph{et al.}~\cite{porzi2015predicting} identified the visual elements that contributed to an image's  perceived level of safety.

But to scale the study of urban perception to multiple cities, and to high spatial resolutions, researchers begun developing computer vision methods to score millions of images \cite{naik2014streetscore}. These computer vision approaches build on research predicting human perception from visual data~\cite{grabner2013visual,isola2011understanding,sartori2015affective} and on research analyzing visual streetscapes for city understanding~\cite{arandjelovic2015netvlad,arietta2014city,doersch2015makes,glaeser2015big,khosla2014looking,zhou2014learning}. The latter of these two lines of research has been fueled by the widespread availability of geolocated image data, such as Google Street View maps or city snapshots publicly shared on social networks (e.g. Flickr, Instagram)~\cite{arandjelovic2015netvlad,sivic2014urban, zhou2014learning}. Using geotagged image data Doersch \emph{et al.}~\cite{doersch2015makes} showed that geographically representative visual elements, like architecture styles, can be automatically discovered from Street View Imagery. In a dynamic study, Naik \emph{et al.}~\cite{naik2015people} used computer vision and images from different time periods to measure urban change, and to study the factors that contribute to neighborhood improvement. Computer vision methods have also been used to show that a city's visual attributes work as proxy of social and economic characteristics, such as crime rates and proximity to local businesses~\cite{arietta2014city,khosla2014looking}, or census characteristics, such as income and inequality~\cite{naik2016cities}. 

This new wave of research has benefited from advances in deep learning, which have been used not only to measure appearance, but also for place recognition.  Zhou \emph{et al.}~\cite{zhou2014learning} introduced Places205 Dataset, a large data collection gathering more than 7 million labeled pictures of scenes. They achieve state-of-the-art classification results by training deep Convolutional Neural Networks (CNNs). 
More recently, Arandjelovi{\'c} \emph{et al.}~\cite{arandjelovic2015netvlad} introduce NetVLAD, a modified CNN architecture able to address large scale visual place recognition. In this work, we build on top of recent work in place recognition~\cite{arandjelovic2015netvlad,zhou2014learning} to fine-tune a deep CNN architecture and show experimentally that under scarce training data, sample augmenting helps achieve state-of-the-art results on safety prediction from streescape images.  

\section{Datasets}
Next, we describe the datasets used to estimate urban appearance, and human activity.

\subsection{Urban appearance data}

We use urban appearance data from the Place Pulse dataset\footnote{\label{footnote:PP}\url{http://pulse.media.mit.edu}}. Place Pulse is a large,
crowdsourcing project on human perception of cities. 
The data collection is designed as an online game, where participants are shown two images of streetscapes 
and are asked to choose one image in response to an evaluative question such as: Which place looks safer? Or: Which place looks more lively? A score is later computed for each image using the TrueSkill~\cite{herbrich2006trueskill} algorithm. 

Place Pulse begun as Place Pulse 1.0 (PP1), which scored 4,109 images from two US cities (New York City and Boston) and two European cities (Linz and Salzburg), and was launched publicly in 2011. PP1 scored images across three evaluative dimensions: \emph{Safety}, \emph{Upper-Class} and \emph{Unique}. 

The current version of Place Pulse (Place Pulse 2.0), launched publicly in July 2013, extended the data collection effort to 56 cities from all continents (except Antartica) and to six evaluative questions: \emph{Safety}, \emph{Wealthy}, \emph{Boring}, \emph{Lively}, \emph{Depressing} and \emph{Beautiful}. For more details, please see Dubey \emph{et al.}~\cite{dubey16eccv}

Here we use data from the Place Pulse 2.0 dataset for Milan and Rome (PP2-I). The data includes 3,897 images that received about 25,000 evaluations for their perception of safety\footnote{We include in this sum only pairwise comparisons involving either two Italian cities or one Italian vs one non-Italian city} - corresponding to an average of 7.6 clicks per image.\\

\subsection{Mobile phone activity data}

To proxy human activity we use mobile phone billing and operation data. The data records the time of communication, and the radio base station that handled it, for various types of communication (e.g. incoming calls, Internet, outgoing SMS). The natural coarsening of this data are grid cells with an area that is proportional to the underlying coverage area of the radio base stations. The cell size is $\sim 300 \times 300$~\si{\m} in the city centre of Milan and it increases up to $\sim 2,300 \times 2,300$~\si{\m} in the peripheral part of the cities, where few customers are served per unit of area. For each grid cell, we count the number of people who made or received a call on an hourly basis, broken down by gender (number of males/females) and age.

Call records were provided by Telecom Italia Mobile (TIM), which is the largest mobile operator in Italy with a market share of $34\%$\footnote{\url{http://bit.ly/1LtNrFY}}.
Our data are aggregated every 60 minutes, and includes both TIM customers and roaming customers in Milan and Rome, and covers the time ranging from February to June 2015.

We note that our data cannot distinguish between pedestrians and people using their phones in their homes, so our measures of activity proxy the number of TIM customers in an area, but not necessarily in the street.

\section{Methodology}

This paper focuses on investigating the relationship between the appearance of safety and the activity or liveliness of neighborhoods in Rome and Milan. To achieve this goal we estimate the appearance of safety for different districts of the city by spatially aggregating the safety scores obtained from the images within these areas, and then, observe people's activity using mobile phone data. To achieve enough coverage, we densify our maps by collecting additional images and scoring them using computer vision. 
Finally, we spatially aggregate scores including census tracts for the 2011 Italian census. In the reminder of this section we explain the methodology used to score images and to measure activity.

\subsection{Measures of Urban Appearance}

\subsubsection{Safety Perception from Visual Data}
\label{subsec:predicting}
We use deep Convolutional Neural Networks (CNNs) as our model for predicting safety perception from streetscape imagery.
We base this choice on the recent success of CNNs in various computer vision problems (especially object classification, object detection and scene recognition~\cite{krizhevsky2012imagenet, oquab2014learning, simonyan2014very}). We show that fine-tuning a CNN on predicting the level of safety along with data augmentation leads to improved performance when compared to recent work on the same task~\cite{naik2014streetscore, ordonez2014learning}. 

Since we have a limited amount of training data (\textit{i.e.} a sample of a few thousands), we retrain CNNs trained on related domains (e.g. images captured in urban environments that are likely to show similar visual content). In particular, we fine-tune the well known AlexNet CNN~\cite{krizhevsky2012imagenet}, trained on Places205 Dataset~\cite{zhou2014learning}, also known as \textit{Places205-AlexNet}. This model was trained on 205 categories of scenes (many of which capture different areas from cities such as office buildings, churches, residential neighborhoods, shops, etc.) summing up around 2.5 million images. By retraining a CNN previously trained in a similar domain we transfer some of the knowledge contained in this network. In our case, we find that the resulting network can accurately predict the appearance of safety, as we show in the validation section (see section 4.1.2). 

To increase the generalization ability of the trained CNN, we adopt data augmentation by cropping all the images used during training and testing. We find this particularly suitable for our scenario where no constraints regarding image alignment are imposed. Specifically, for every image in the training set, we generate \textit{n} crops by randomly assigning values to the coordinates of the top left and bottom right cutting points, respectively. We control the size of the crops by bounding the coordinates of the cutting points to ranges proportional with the image size. Formally, given an image $I$ of size $W\times H\times 3$, we generate points $P_1(x_1, y_1)$ and $P_2(x_2, y_2)$ such that the quantities: $x_1/W, y_1/H, 1-x_2/W$ and $1- y_2/H$ are bounded by $k_1$ and $k_2$, respectively. We empirically set $k_1$ to $0.05$, $k_2$ to $0.2$ and $n$ to $30$ in all our experiments. Additional constraints to control diversity of the crops could be implemented, such as monitoring the \textit{intersection over the reunion (IoU)} between pairs of crops and/or original image. During testing, we average all predictions belonging to the crops of the same test sample. The safety scores obtained at image level are then aggregated into city's districts.

Part of the standard pre-processing steps, all the images are subject to scaling to $227 \times 227$ pixels and mean image subtraction. The task is modeled as a regression problem, where the goal is to minimize the $l_2$ loss between the sample labels and the model predictions. As labels, we refer to the Trueskill scores computed in~\cite{naik2014streetscore}, also used in~\cite{ordonez2014learning}. We fine-tune with \textit{Caffe}~\cite{jia2014caffe} for $10,000$ iterations using a base learning rate of $1e^{-4}$. We note no further decrease in the training loss beyond this limit.

\subsubsection{Validation of computer vision models}
We validate the fine-tuned CNN on the two US cities from PP1 (New York and Boston), following training and evaluation protocols from~\cite{naik2014streetscore} and~\cite{ordonez2014learning}, respectively. In the first case, we perform a 5-fold cross-validation on the 2920 US images and report the average $R^2$ measure, after scaling all the scores to the interval $[0, 10]$. We obtain an average $R^2$ of $62.2\%$ a $9.5\%$ relative improvement over the best result reported in~\cite{naik2014streetscore} and $4.8\%$ over the case of not using data augmentation during testing. In the second case, the source (train) and target (test) domains are populated by alternating US cities (e.g. NY - Boston, NY - NY, Boston - Boston, Boston - NY). As performance measure, the authors report Pearson correlation between the predicted regression values and the label scores. Table~\ref{tab:corr_ordonez} shows the comparison results, where we consistently outperform~\cite{ordonez2014learning} in all four combinations. We also observe how the difficulty of the task (for different pairs) modulates the predicting performance of both sets of models in the same way (e.g. training and testing on Boston seems to be the easiest for both our models and the ones from~\cite{ordonez2014learning}).  

\begin{table}[ht]
    \centering
    \small
    \begin{tabularx}{\columnwidth}{@{}Xcc@{}}
        \toprule
        \textbf{Model type} & \textbf{Best from~\cite{ordonez2014learning}} & \textbf{Our result} \\ \midrule
        NY - NY  & 0.687 & 0.718  \\
       Boston - Boston    & 0.718 & 0.744  \\ 
       NY - Boston   & 0.701 & 0.734  \\
       Boston - NY  &   0.636  &  0.693    \\
         \bottomrule
    \end{tabularx}
    \caption{Performance on the estimated level of safety (comparison with~\cite{ordonez2014learning}). For all cases, we report Pearson correlations with $\mathbf{p<0.001}$.}
    \label{tab:corr_ordonez}
\end{table}

\subsubsection{Densification of Appearance data using Computer Vision}
\label{sec:densifying}

Since the distribution of the annotated images from PP2-I is sparse, on average 6.7 images/\si{\square\km},
we retrieve additional images for the cities of Rome and Milan by densely sampling geo-referenced images from Google Street View.
Data densification for the analysis of urban landscape has been used in previous works for generating high resolution maps of city perception~\cite{naik2014streetscore,ordonez2014learning}.
First, we generate a grid of points inside the area we want to cover. 
The granularity we choose is 100 points/\si{\square\km}. To better represent the safety perception of the location, we retrieve four images from each location, which have 90 degrees horizontal field of view and different headings (north, east, south and west). By doing this we cover 360 degrees from each location, thus getting less biased safety perception of an area by averaging predicted safety scores of these four images. 
We developed the script which iterates through all the points and used Google Street View API to obtain four images for each location. 
The script discards locations where no images are available. Using this method we obtained 83,203 images for Rome and 74,815 images for Milan. 

\subsubsection{Validation of Densification}
Table~\ref{tab:corr_italy_placepulse} reports the performances on predicting safety perception for the city of Rome and Milan.
We evaluate the correlation between the aggregated original scores and (i) the predictions over the images from PP2-I and (ii) the predictions over the images from the densified dataset.
In general, a slight decrease in performance is observed in the second case. We can attribute this loss in performance to the fact that images from PP2-I were retrieved in 2010, thus a variation in the urban appearance may have occurred within this time lapse.
For example, the Expo Milano 2015 - a Universal Exposition - was held from May to October 2015, and Milan underwent a profound (visual) change in its northwest area to prepare for this event.

\begin{table}[ht]
    \centering
    \small
    \begin{tabularx}{\columnwidth}{@{}Xcc@{}}
        \toprule
        \textbf{City} & \textbf{PP2-I} & \textbf{Densified}\\ \midrule
        Milan          & 0.621  & 0.488  \\
        Rome          & 0.635  & 0.548 \\
         \bottomrule
    \end{tabularx}
    \caption{Performance on the estimated safety perception level for each city. All values are statistically significant, with $\mathbf{p<0.001}$.}
    \label{tab:corr_italy_placepulse}
\end{table}

\begin{figure*}[ht!]
\centering
\includegraphics[width=0.8\textwidth]{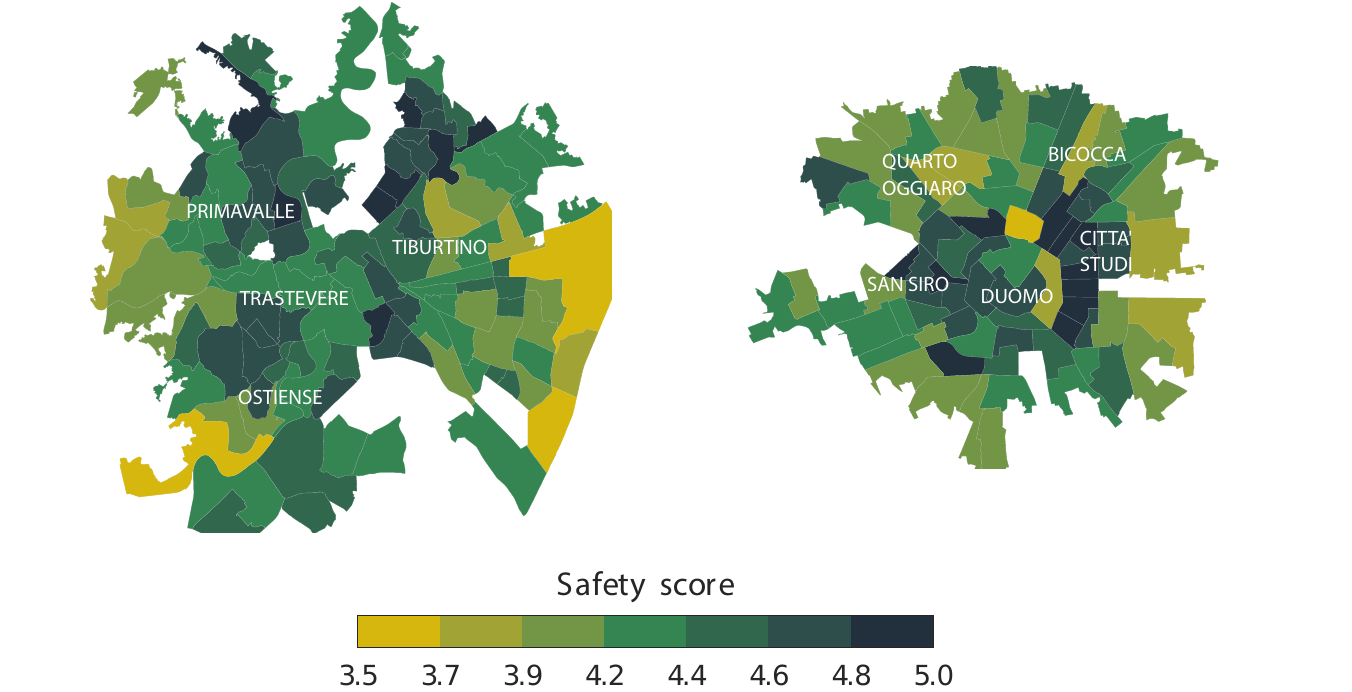}
\caption{Spatial distribution of perceived safety in each district of Rome and Milan.} 
\label{fig:teaser}
\end{figure*}

\textbf{Predicting safety on Milan and Rome:}
We are interested in finding a good candidate model for labeling the densely sampled Street View images from Milan and Rome. We experimented with several model choices (including training on PP2-I) and discovered that, surprisingly, using PP1 for training yields the best correlation value for the two Italian cities. We attribute this result to the much inferior average number of votes per image (around 7.6 for PP2-I, compared to around 90 for PP1). For the rest of the experiments, we use only the model trained on PP1 for safety prediction.
In \Cref{fig:teaser} we visually report the spatial distribution of the safety prediction for Milan and Rome.

\subsection{Metrics for Urban Liveliness or Activity}

Next, we define the metrics we use as proxies for an area's level of activity, or liveliness. 
Here we study only in urban areas where more than 50\% of the surface is not composed by farmlands or forests.
We measure activity for four populations, all people, females, people younger than 30, and people older than 50.

Formally, we measure the density of all people in district $i$ as:

\begin{align}
    R_p(i, 24h) &= \frac{|people_{i, 24h}|}{area_i} 
\label{eq:R_p}
\end{align}

Next, we measure the fraction of females in a district $i$ as: 

\begin{align}
    R_{f}(i, 24h) &= \frac{|females_{i, 24h}|}{|people_{i, 24h}|}
\label{eq:R_f}
\end{align}

Additionally, we measure the population of people below 30 and above 50 as:
\begin{align}
    R_{<30}(i, 24h) &= \frac{|people(<30)_{i, 24h}|}{|people_{i, 24h}|}
\label{eq:R_y}\\
    R_{>50}(i, 24h) &= \frac{|people(>50)_{i, 24h}|}{|people_{i, 24h}|}
\label{eq:R_o}
\end{align}.

\subsection{Spatial Regression}

We test the connection between urban appearance of safety and activity using spatially corrected Ordinary Least Squares (OLS) regressions. Since we are dealing with spatial variables, OLS residuals are assumed not to be spatially auto-correlated; otherwise the regression model is said to be misspecified.
Thus, we use the Griffith filtering~\cite{tiefelsdorf2007semiparametric} which extracts a set of orthogonal and uncorrelated eigenvectors from the expression:
\begin{equation}
\label{eq:grif}
    (I - \frac{11^T}{n})W(I -\frac{11^T}{n})
\end{equation}
derived from the spatial auto-correlation Moran's I numerator, where $I$ is a  ($n \times n$) identity matrix, $1$ is a $n \times 1$ vector containing only 1's and $W$ is a ($n \times n$) spatial weight matrix based on topological adjacency, so-called Queen criterion: if two areas share a boundary or a vertex, the entity of the spatial weight matrix is coded as 1, and otherwise, 0.
The eigenvectors obtained can be employed in a multivariate regression to account for spatial auto-correlation. 
However, it is clear that employing all $n$ eigenvectors in a regression framework is not desirable for reasons of model parsimony.
Thus, a subset of eigenvectors are selected in a step-wise fashion so as to minimize the sequential residual spatial correlation (Moran's I) values~\cite{tiefelsdorf2007semiparametric}.
The final subset of candidate eigenvectors represents the \emph{spatial filter} for the variable analysed.
Before applying the regression model, the data were Z-score scaled.

\section{Results}

After describing our methods to measure urban appearance and neighborhood activity we test whether the appearance of safety and the activity of neighborhoods is correlated. To test for this correlation we merge our data with information from the Italian census so we can control for other sources that intuitively correlate with neighborhood activity: population density (residential), density of employment (which should also proxy pedestrian density during the day), distance to city centre, and a deprivation index (to control for poverty in the neighborhood).

We begin by looking simply at the correlation between the number of people per unit of area observed in our mobile phone dataset and the appearance of safety in a neighborhood while controlling for population density, employee density, deprivation, and distance to centre. 

\Cref{tablepeople} shows the result of a spatially corrected multivariate regression with the number of people per unit of area  measured using the density of all people as the dependent variable. Not surprisingly, the strongest correlate of the number of people present per unit of area is employee density, and the number of people per unit area decreases with distance to the center. Yet, despite the strong effect of the other control variables, the appearance of safety is significantly and positively correlated with the number of people present in a neighborhood per unit area. 

\begin{table}[ht!]
    \centering
    \small
    \begin{threeparttable}
    \begin{tabularx}{\columnwidth}{@{}Xr@{}}
        \textbf{Presence of people (\ref{eq:R_p})\tnote{l}} \\ \midrule
        Population density\tnote{l} 		& 0.155**  \\
        Employees density\tnote{l}   & 0.328** \\
        Deprivation        & -0.022  \\
        Distance centre & -0.257** \\
        Safety appearence & 0.105** \\
        \midrule
        Spatial Eigenvectors & 11  \\
        Adj-$R^2$ & 0.91  \\
        Moran's I (p-value)  & 0.07 (0.08)  \\
         \bottomrule
    \end{tabularx}
    \begin{tablenotes}
        \item [l] $\log$ transformed variable.
        \end{tablenotes}
    \end{threeparttable}
    \caption{OLS regression model between presence of people and safety perception. The $\mathbf{\beta}$ coefficients are reported in the table. *$\mathbf{p<0.01}$, **$\mathbf{p<0.001}$.}
    \label{tablepeople}
\end{table}

Next, we look at the fraction of females present in an area. Looking at the population of females separately is motivated by empirical research showing that women are twice as likely as men to report feeling unsafe~\cite{wekerle1995safe}, even though they have a much smaller risk of being victimized~\cite{pantazis2000fear, taylor1986testing}. This suggest that the presence of women in a neighborhood should be more strongly affected by its appearance of safety than the presence of men. In fact, Felson and Clarke~\cite{felson1998opportunity} suggest that a high ratio of women in the street is a positive sign towards urban safety, as they act as ``crime detractors," in agreement with Jacobs' natural surveillance hypothesis. These theories would suggest that the ratio of females is lower in places perceived as unsafe.

\Cref{tab:women} looks at the ratio of females in the population observed in our cell phone data as the dependent variable, finding that the appearance of safety is highly significant and positive. In fact, the coefficient is roughly twice that observed for the general population. 

\begin{table}[ht!]
    \centering
    \small
    \begin{threeparttable}
     \begin{tabularx}{\columnwidth}{@{}Xr@{}}
        \textbf{Presence of women (\ref{eq:R_f})}  \\ \midrule
        \% of women (residents)\tnote{s} & 0.001\\
        Deprivation        & -0.005\\
        Distance centre & -0.003\\
        Safety perception & 0.020**\\
        \midrule
        Spatial Eigenvectors & 12\\
        Adj-$R^2$ & 0.65\\
        Moran's I (p-value)  & 0.06 (0.11)\\
         \bottomrule
    \end{tabularx}
    \begin{tablenotes}
        \item [s] cube-root transformed variable.
        \end{tablenotes}
    \end{threeparttable}
        \caption{OLS regression model between presence of women and safety perception. The $\mathbf{\beta}$ coefficients are reported in the table. *$\mathbf{p<0.01}$, **$\mathbf{p<0.001}$.}
    \label{tab:women}
\end{table}

We also look at the proportion of people younger than 30 and older than 50 in an area. According to Felson and Clarke~\cite{felson1998opportunity}, a younger population is a predictor of criminal incidents in an area, as they show a higher aggression potential compared to older populations.
Nevertheless, younger people, especially men, show less fear of crime~\cite{hollway1997risk, mark1984fear}. 

\Cref{tab:young} looks at the ratio of people younger than 30 in the population observed in our cell phone data as the dependent variable.
The variable which contributes the most to the correlation is the distance from the city centre, followed by appearance of safety which is highly significant and negative. 
Contrarily, when we look at the ratio of people older than 50 (\Cref{tab:elderly}), we find that the appearance of safety is highly significant and positive. This is in agreement with the theory, showing that older people are more likely to be present in places that appear safe.

\begin{table}[ht!]
    \centering
    \small
    \begin{threeparttable}
  \begin{tabularx}{\columnwidth}{@{}Xr@{}}
        \textbf{Presence of people younger than 30 (\ref{eq:R_y})\tnote{l}}  \\ \midrule
        \% of younger residents\tnote{l} & -0.001 \\
        Deprivation        & 0.032**  \\
        Distance centre & -0.150** \\
        Safety perception & -0.048** \\
        \midrule
        Spatial Eigenvectors & 16  \\
        Adj-$R^2$ & 0.66  \\
        Moran's I (p-value)  & 0.07 (0.09)  \\
         \bottomrule
    \end{tabularx}
    \begin{tablenotes}
        \item [l] log transformed variable.
        \end{tablenotes}
    \end{threeparttable}
    \caption{OLS regression model between presence of younger people and safety perception. The $\mathbf{\beta}$ coefficients are reported in the table. *$\mathbf{p<0.01}$, **$\mathbf{p<0.001}$.}
    \label{tab:young}
\end{table}

\begin{table}[ht!]
    \centering
    \small
    \begin{threeparttable}
  \begin{tabularx}{\columnwidth}{@{}Xr@{}}
        \textbf{Presence of elderly people (\ref{eq:R_o})}\\ \midrule
        \% of elderly residents\tnote{s} & 0.006** \\
        Deprivation        & -0.006**  \\
        Distance centre & 0.006** \\
        Safety perception & 0.017** \\
        \midrule
        Spatial Eigenvectors & 14  \\
        Adj-$R^2$ & 0.64  \\
        Moran's I (p-value)  & 0.07 (0.09)  \\
         \bottomrule
    \end{tabularx}
    \begin{tablenotes}
        \item [s] cube-root transformed variable.
        \end{tablenotes}
    \end{threeparttable}
    \caption{OLS regression model between presence of elderly people and safety perception. The $\mathbf{\beta}$ coefficients are reported in the table. *$\mathbf{p<0.01}$, **$\mathbf{p<0.001}$.}
    \label{tab:elderly}
\end{table}

\mbox{ }

\begin{figure*}[t]
\centering
\frame{\includegraphics[width=0.16\linewidth]{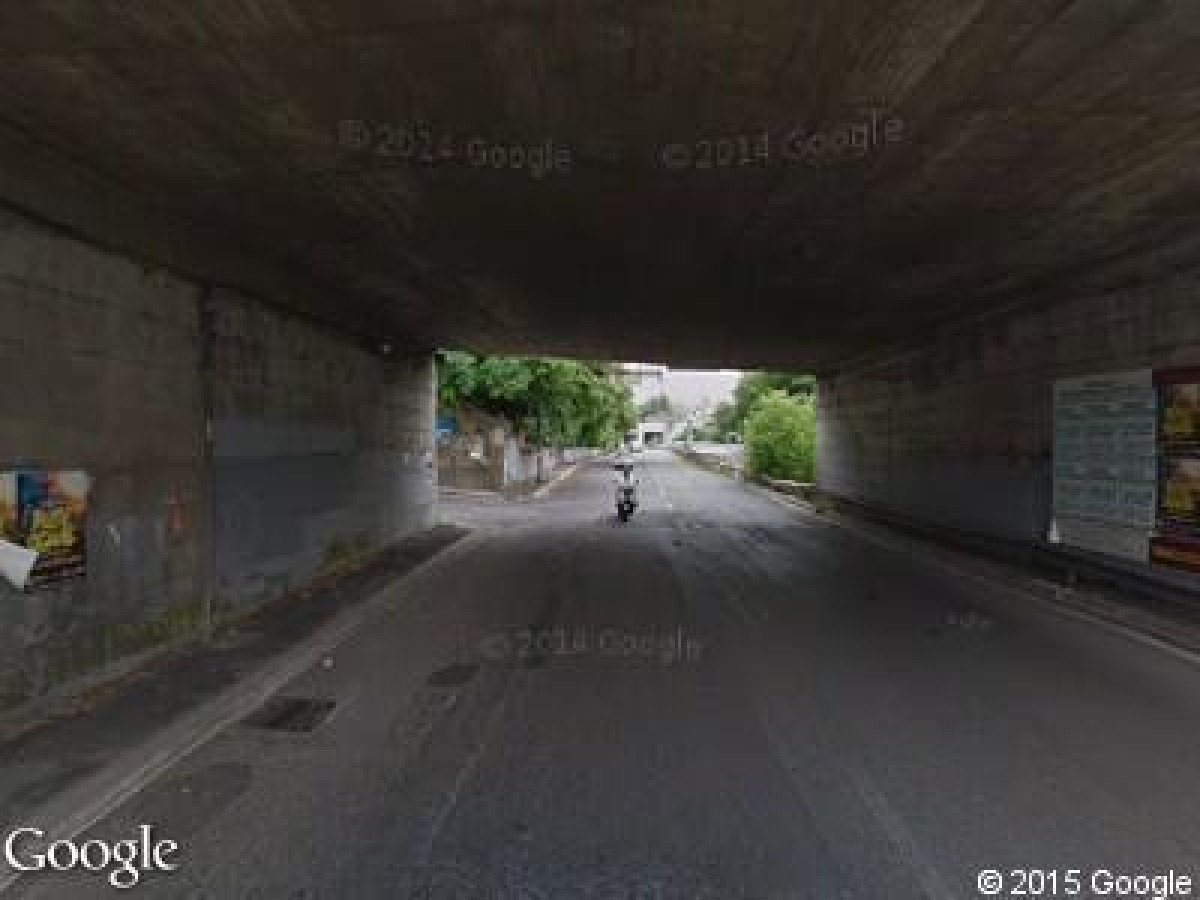}}
\frame{\includegraphics[width=0.16\linewidth]{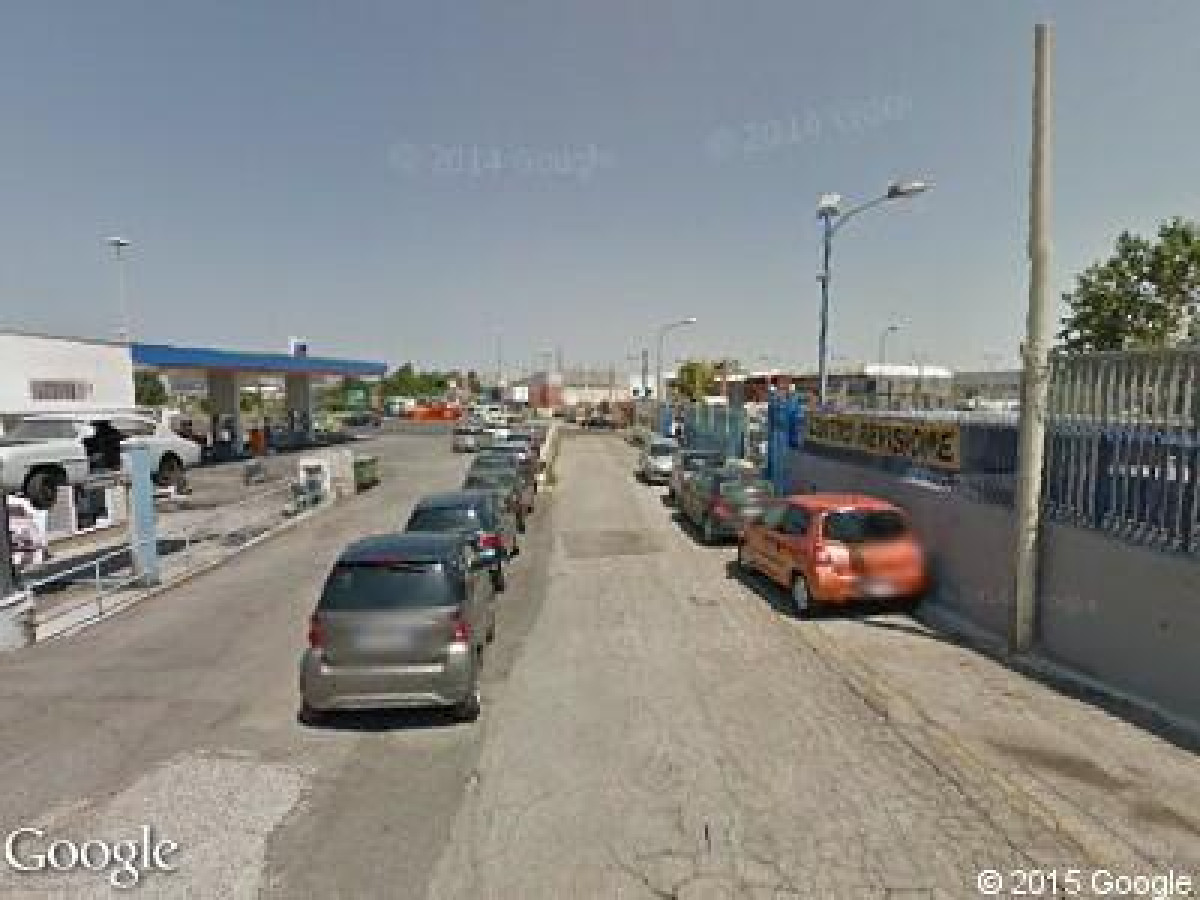}}
\frame{\includegraphics[width=0.16\linewidth]{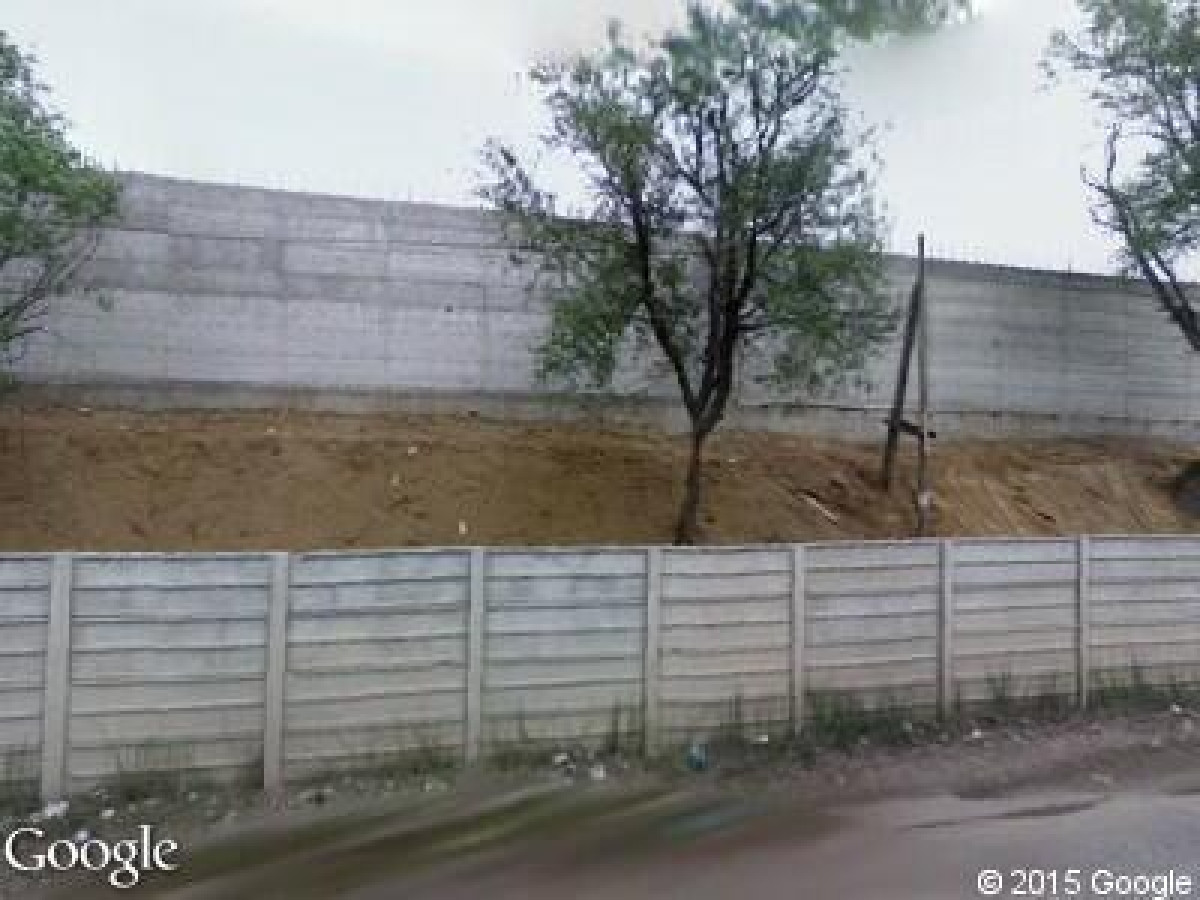}}
\frame{\includegraphics[width=0.16\linewidth]{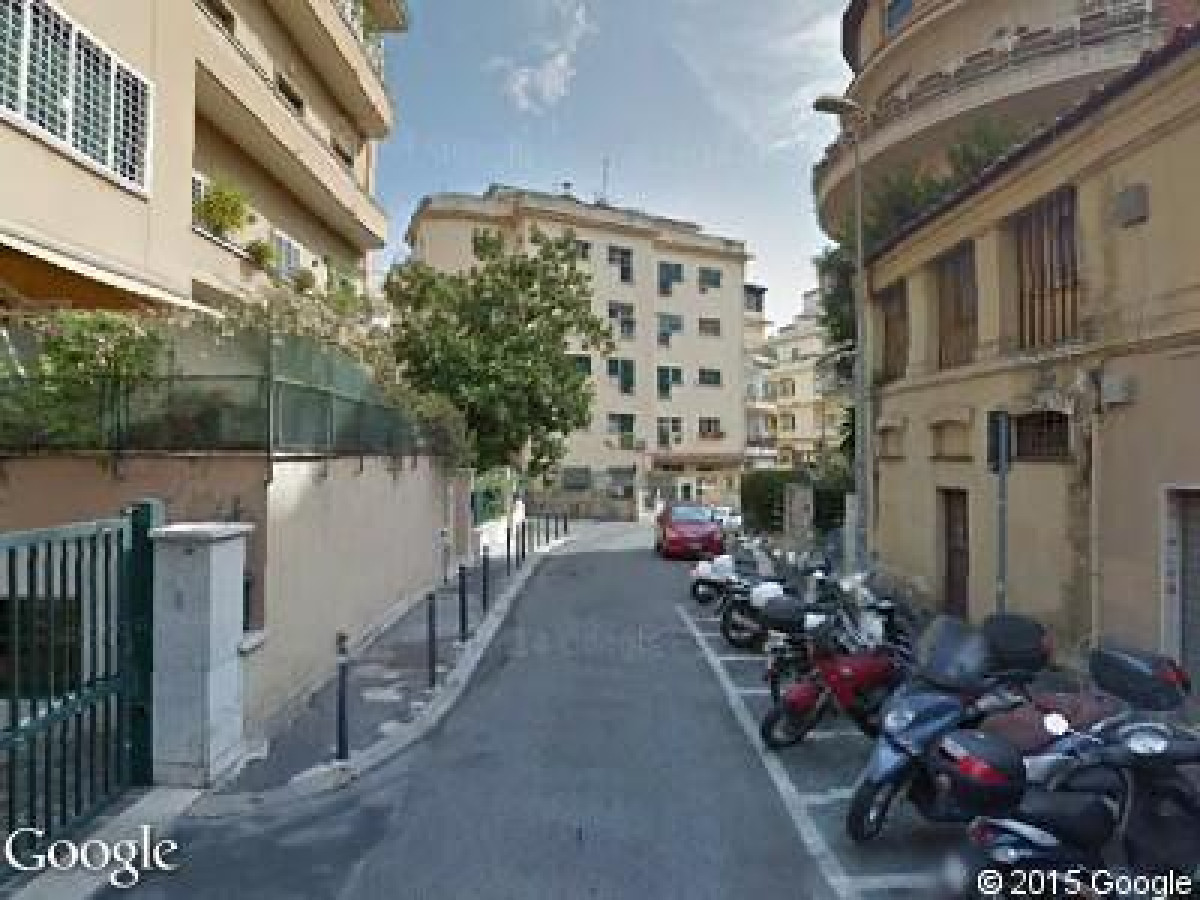}}
\frame{\includegraphics[width=0.16\linewidth]{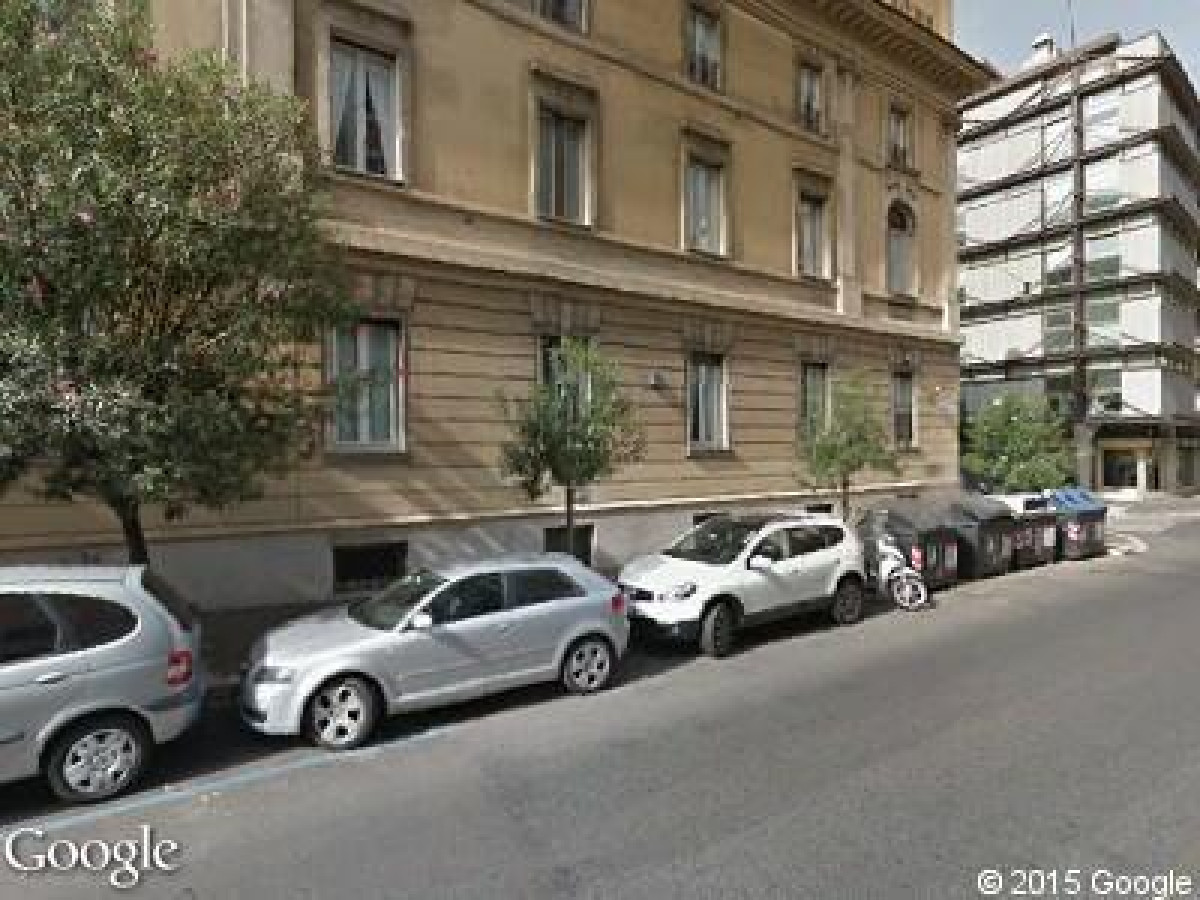}}
\frame{\includegraphics[width=0.16\linewidth]{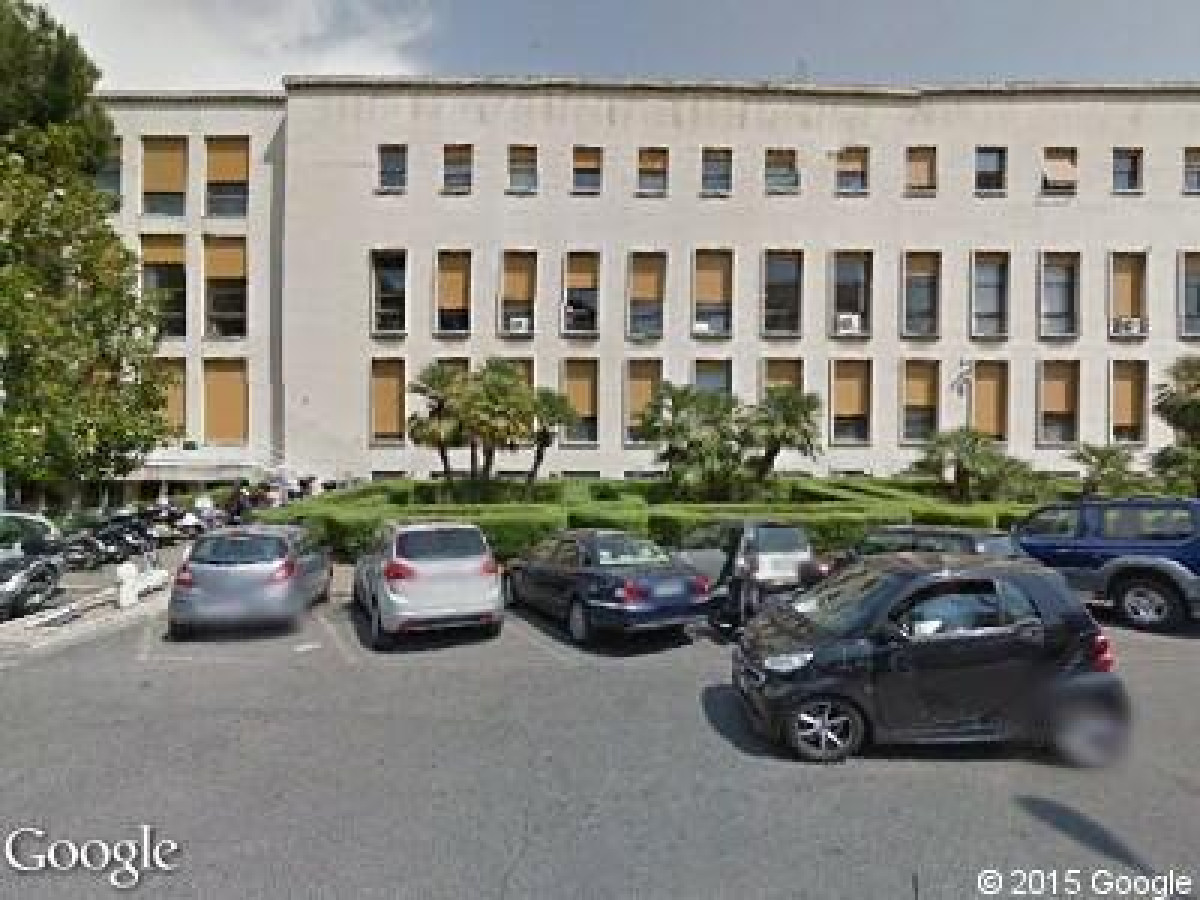}} 
\\
\frame{\includegraphics[width=0.16\linewidth]{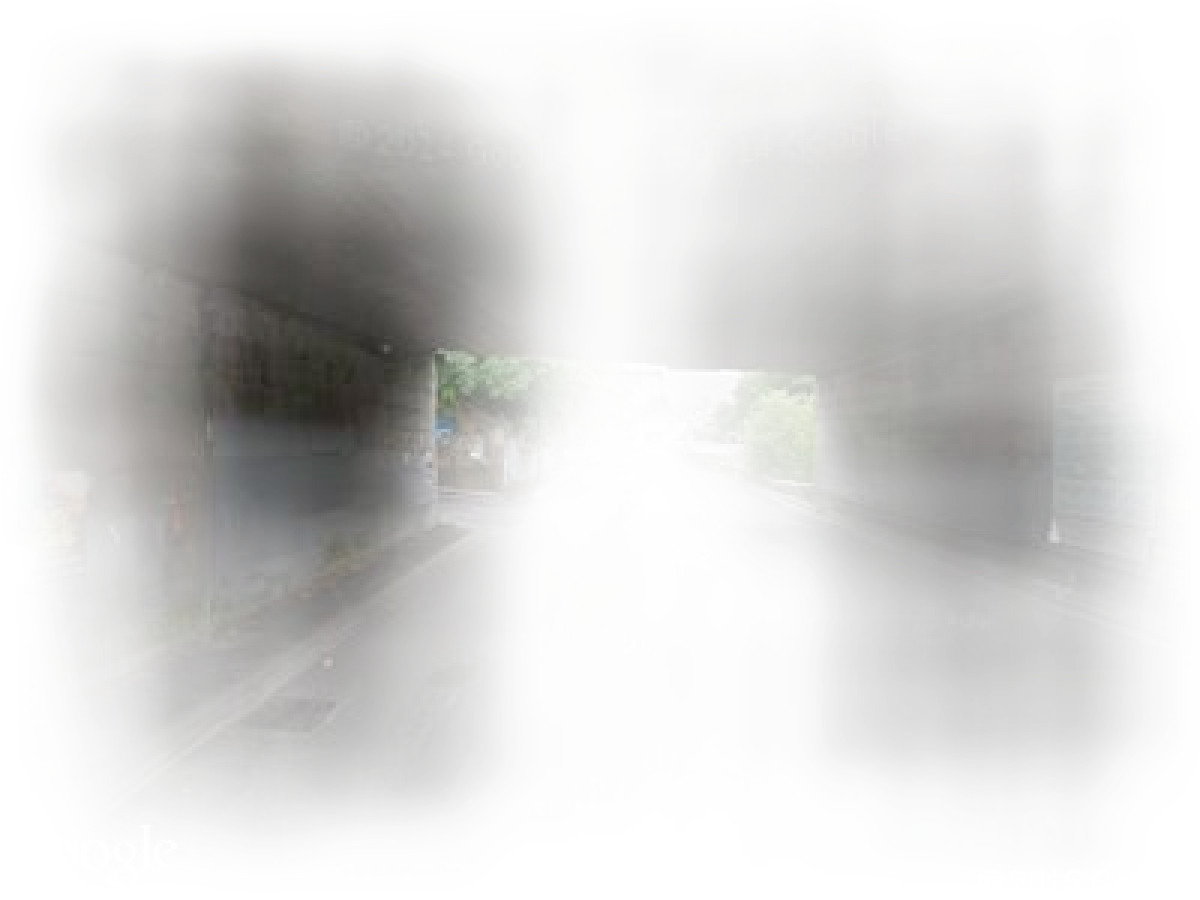}}
\frame{\includegraphics[width=0.16\linewidth]{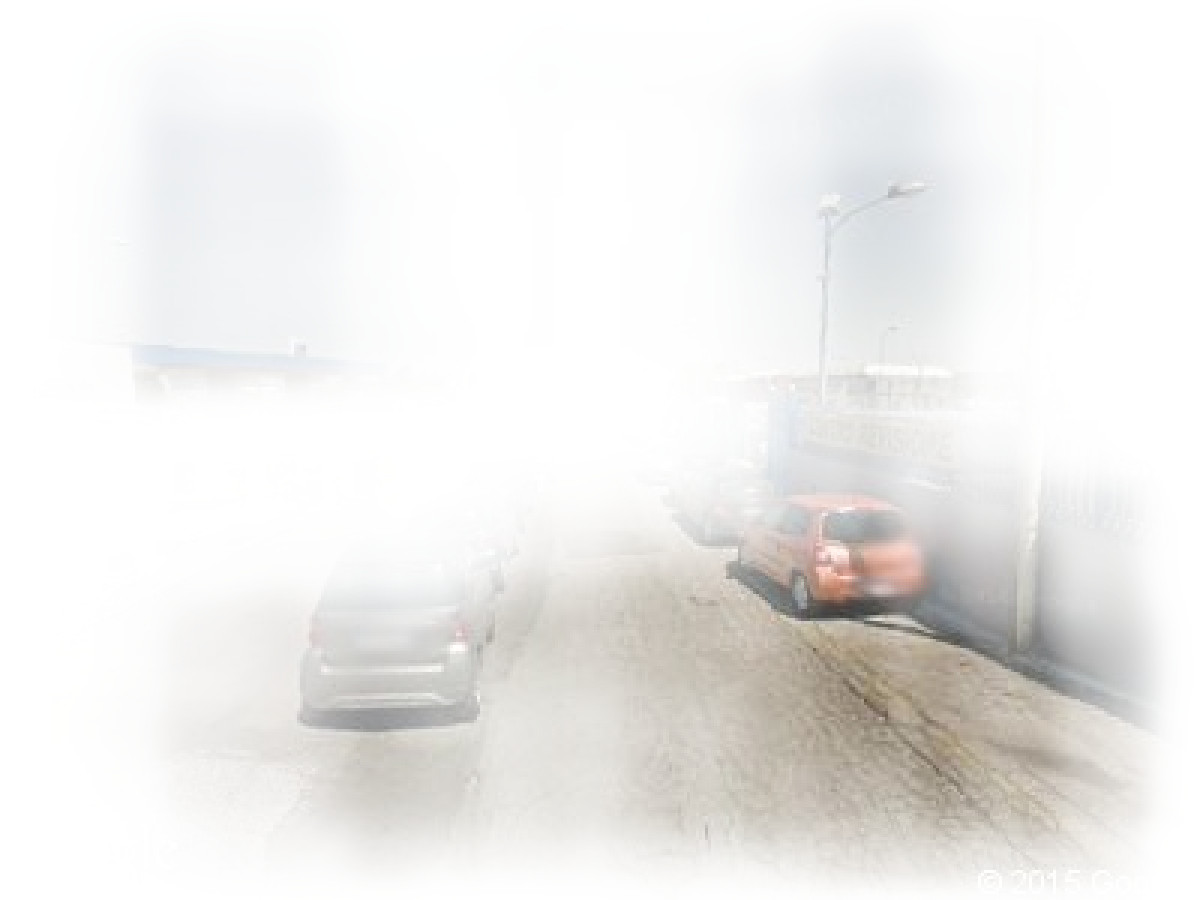}}
\frame{\includegraphics[width=0.16\linewidth]{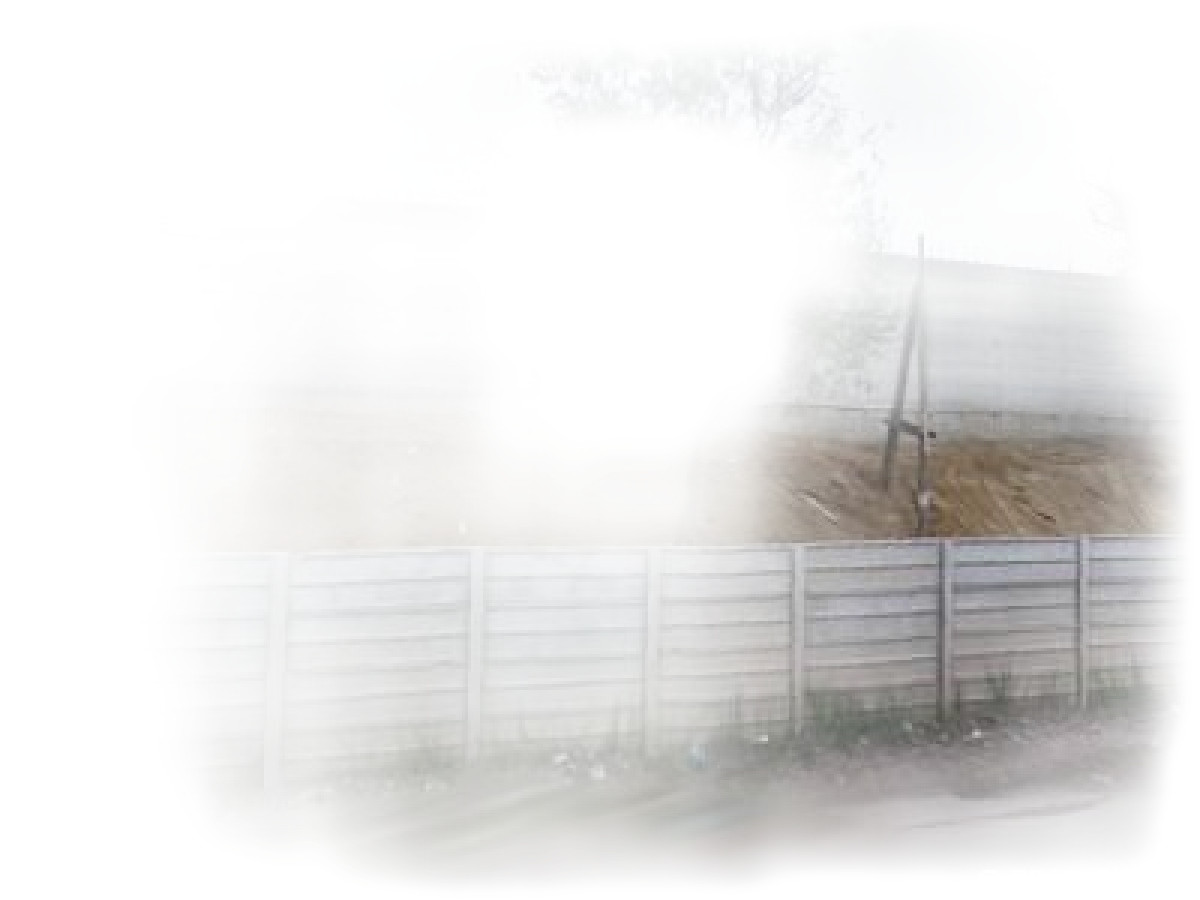}}
\frame{\includegraphics[width=0.16\linewidth]{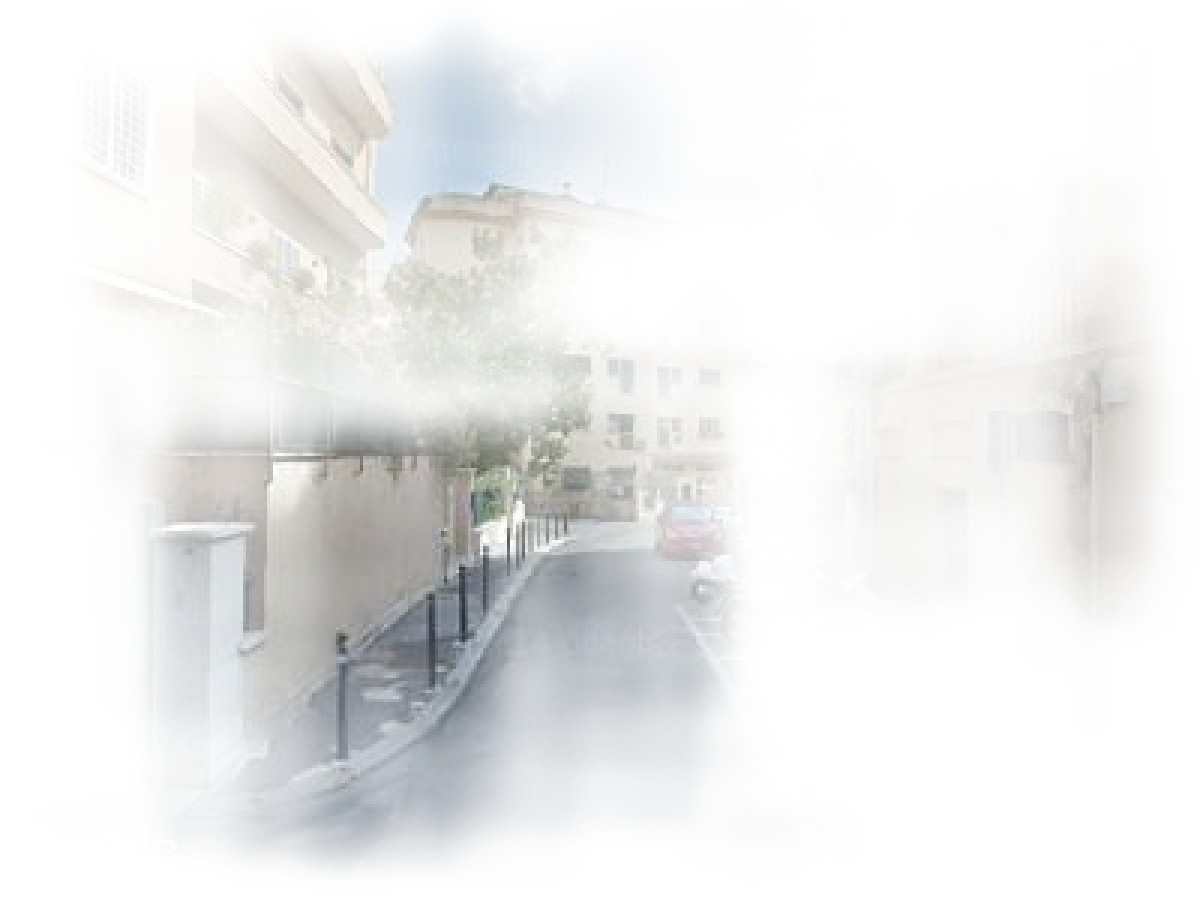}}
\frame{\includegraphics[width=0.16\linewidth]{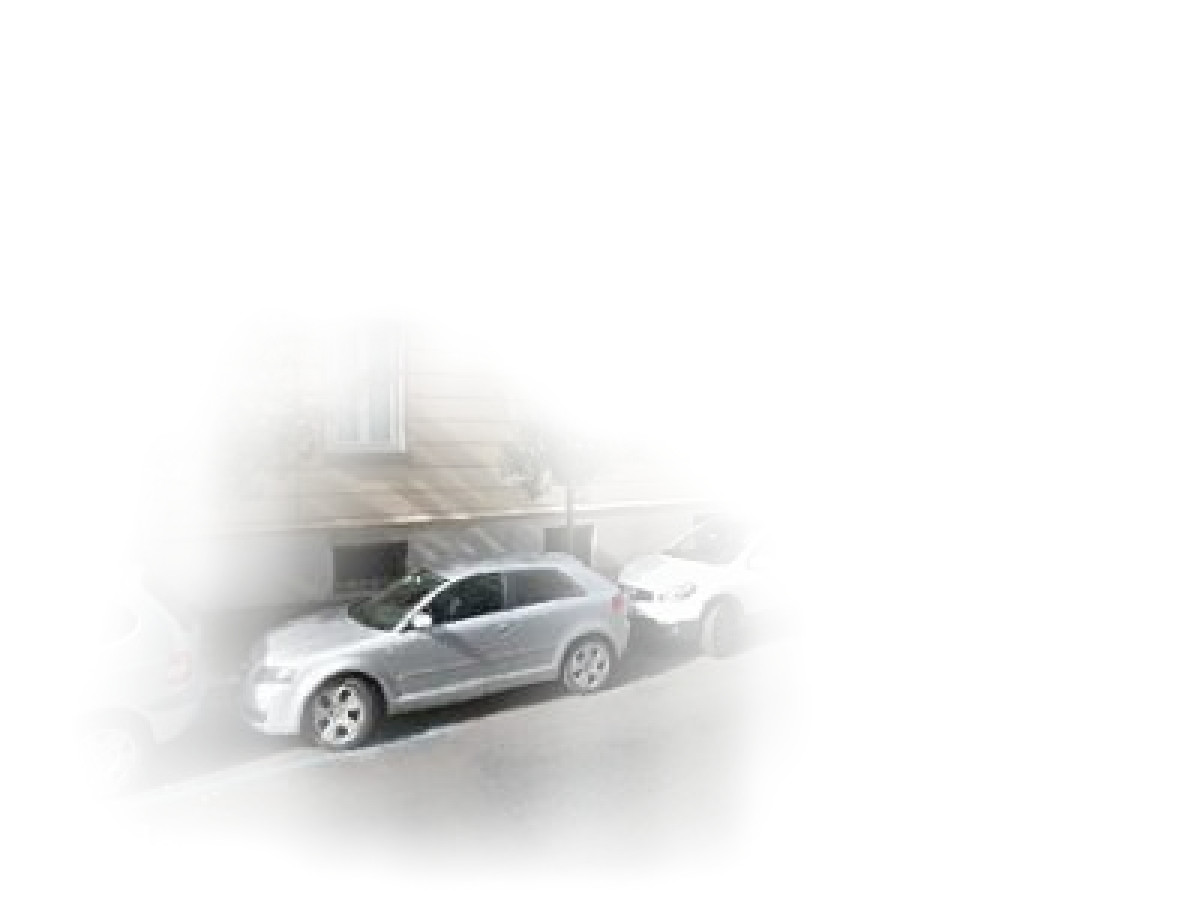}} 
\frame{\includegraphics[width=0.16\linewidth]{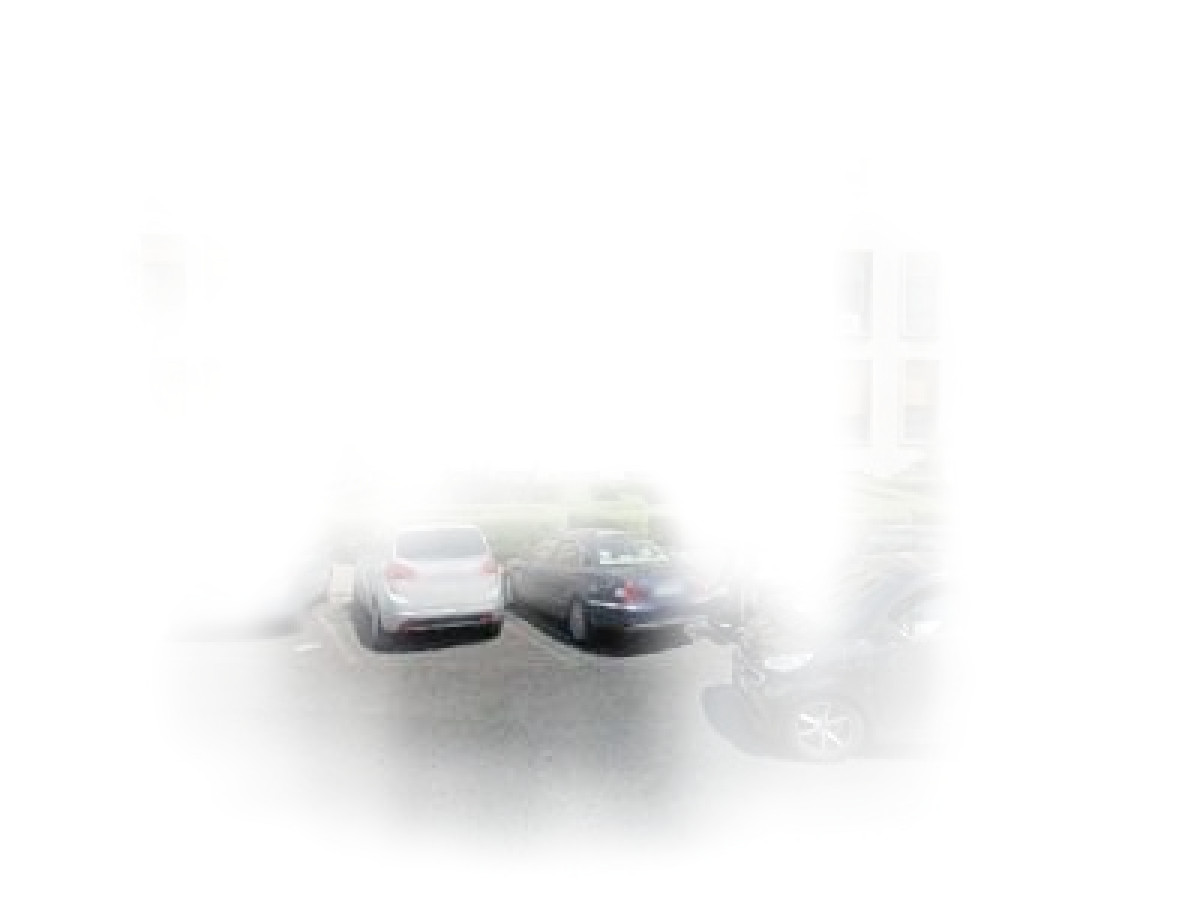}} 
\\
\frame{\includegraphics[width=0.16\linewidth]{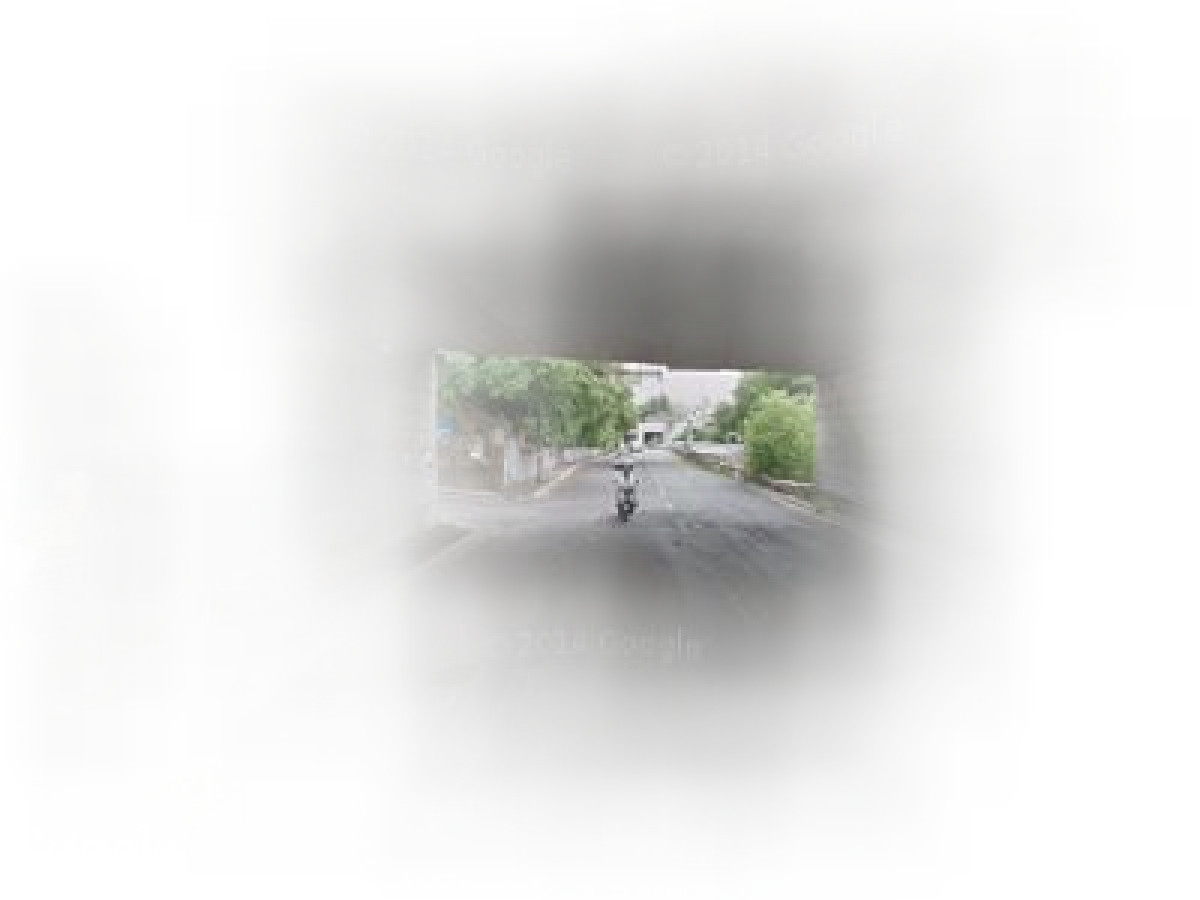}}
\frame{\includegraphics[width=0.16\linewidth]{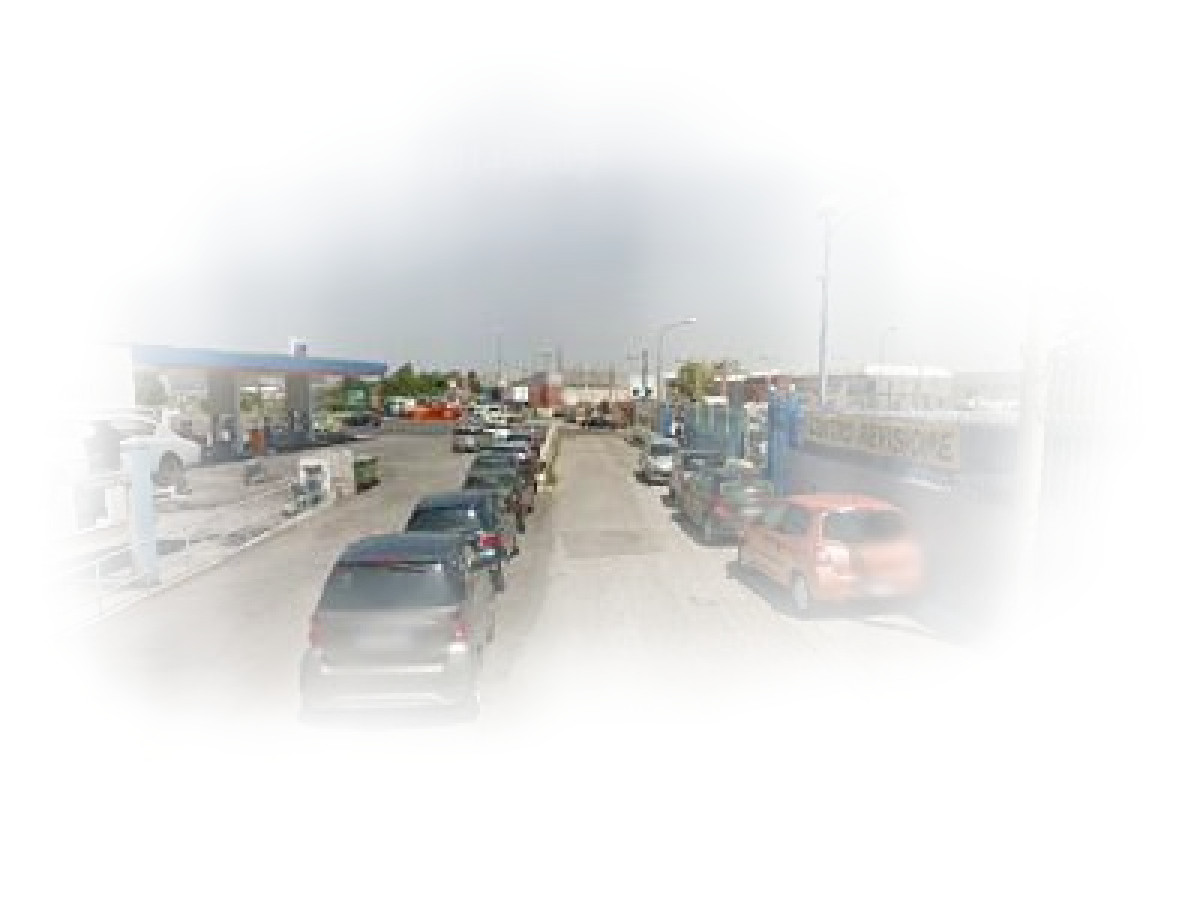}} 
\frame{\includegraphics[width=0.16\linewidth]{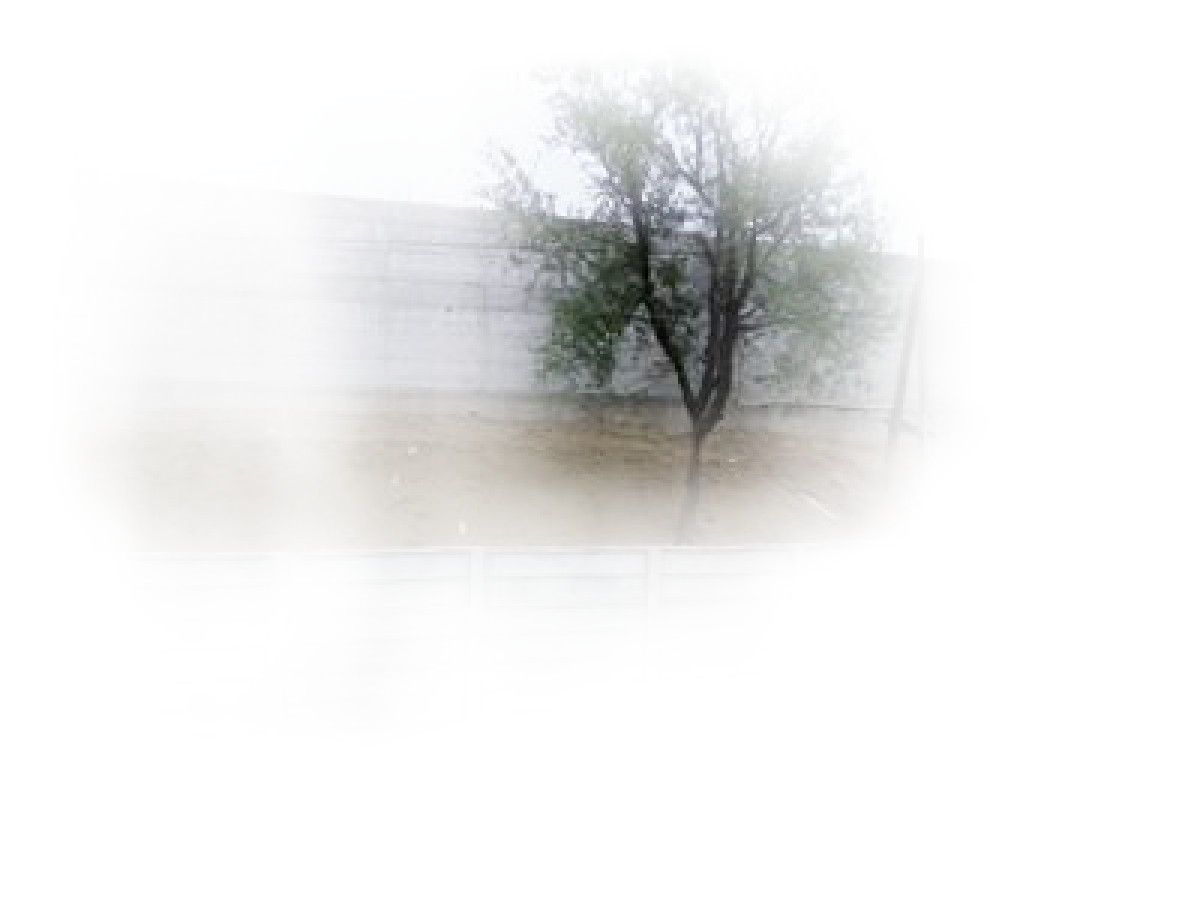}}
\frame{\includegraphics[width=0.16\linewidth]{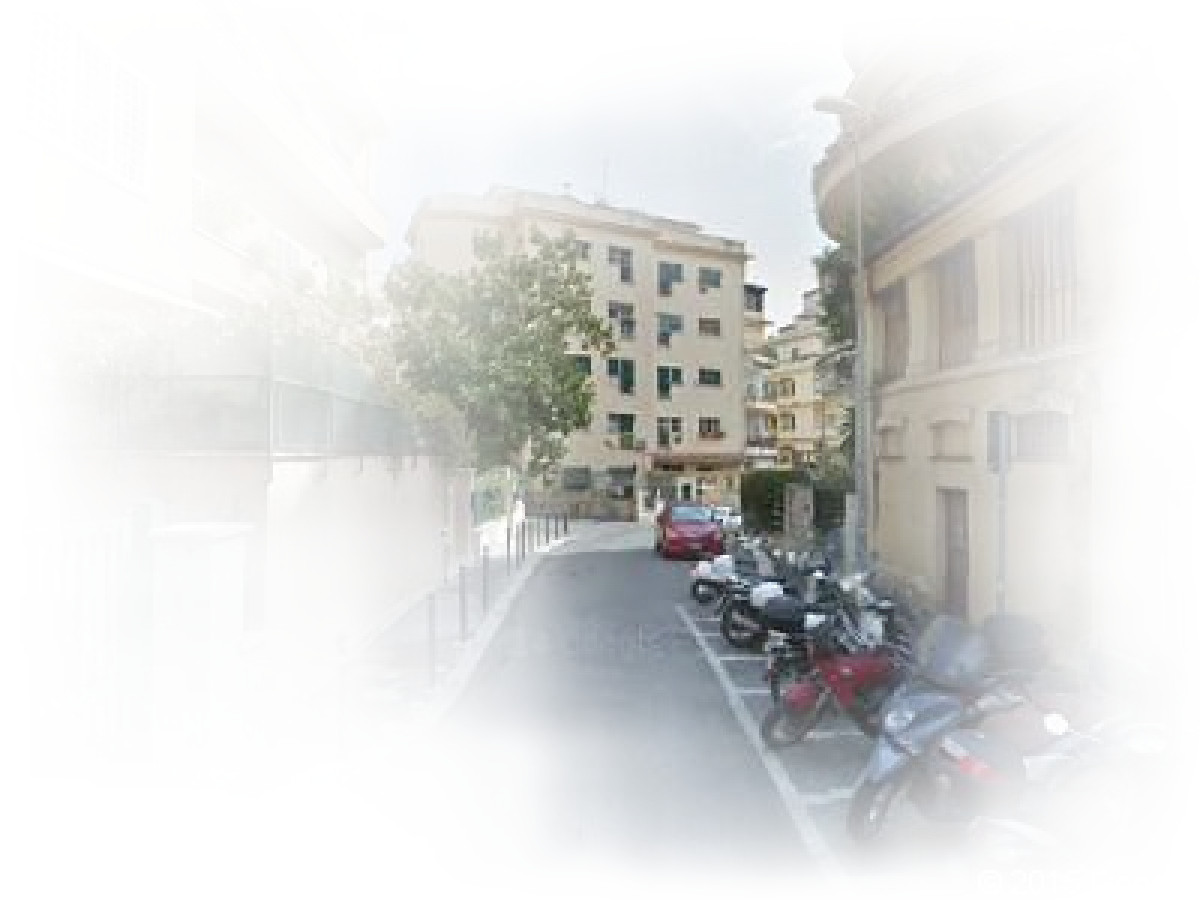}}
\frame{\includegraphics[width=0.16\linewidth]{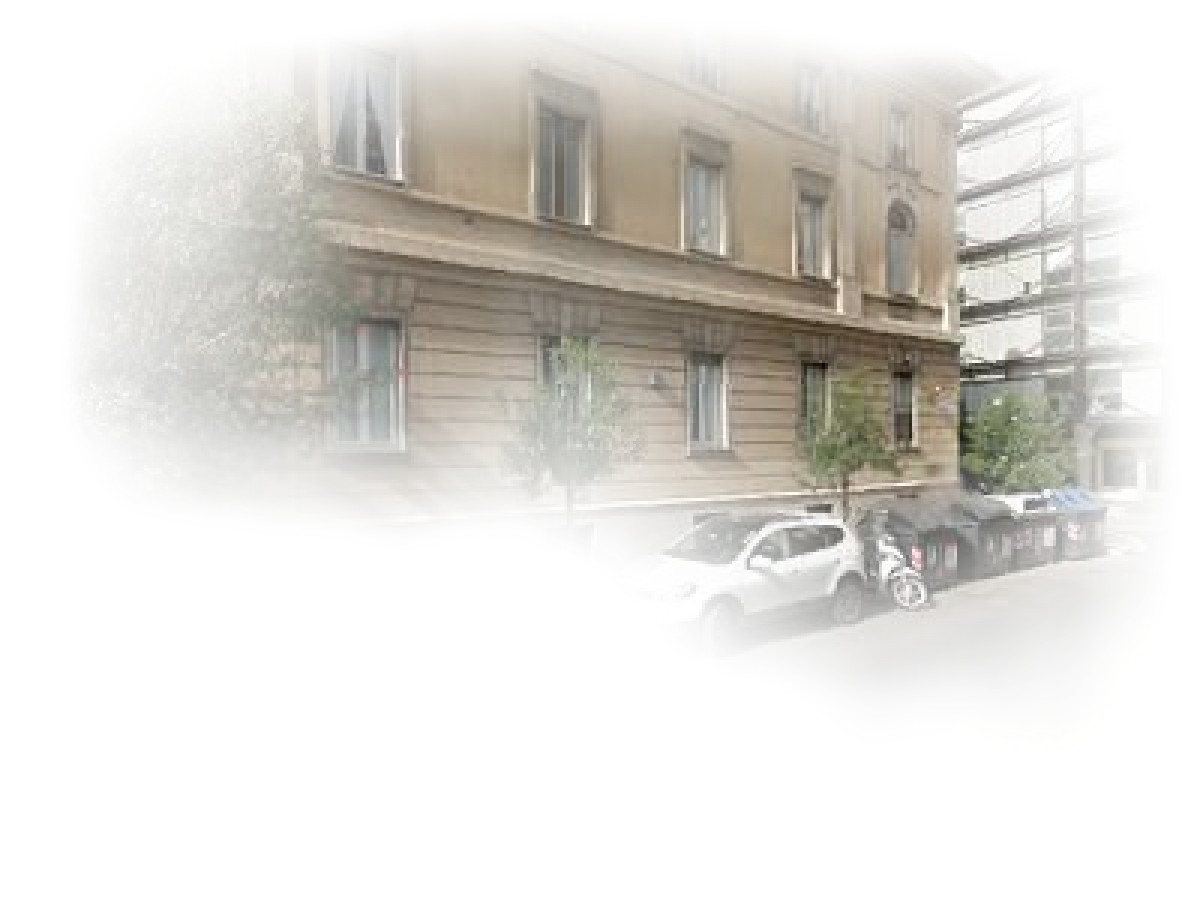}} 
\frame{\includegraphics[width=0.16\linewidth]{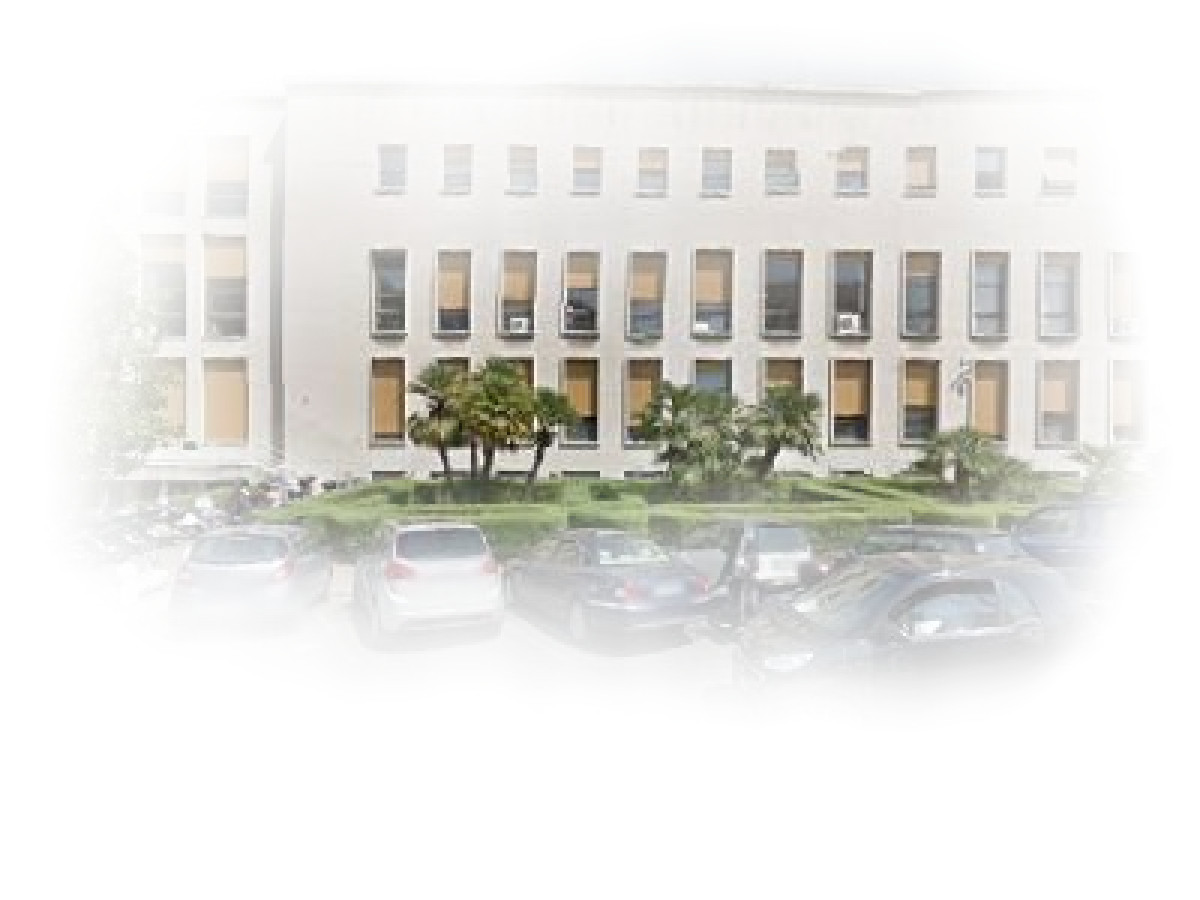}} 
\\
\caption{(Top) Sample images associated to a (left) low and (right) high level of safety and corresponding activation masks: highlighted areas correspond to the ones that mostly contribute to the perception of (center) unsafety and (bottom) safety. }
\label{fig:sampleSafety}
\end{figure*}

\mbox{ } 


\subsection{Visual Attributes Determining Safety}
\label{sec:res:visual}

Finally, we explore the visual attributes of the images that contribute positively, or negatively, to their appearance of safety. To identify these attributes we set up an occlusion sensitivity experiment. In this experiment, inspired by \cite{zeiler2014visualizing}, we randomly generate occluding patches in images and replace them with the average pixel value. For every such altered image, we monitor the effect at the output of the predictor (did the image score higher or lower in its appearance of safety). This allows us to identify patches in an image that contribute positively or negatively to their appearance of safety. 

Figure~\ref{fig:sampleSafety} shows some examples, with the original image on the (top row), followed by the areas that contribute to a low appearance of safety (middle row) and a high appearance of safety (bottom row). The images are sorted from an overall low appearance of safety to a high one. The examples, while illustrative instead of comprehensive, show that street facing windows and greenery tend to to contribute positively to an streetscapes appearance of safety. The positive effect of street facing windows is in agreement with the natural surveillance hypothesis of Jane Jacobs. 

\section{Conclusion}
In this paper we explored the question: ``\emph{Are safer looking neighborhoods more lively?}'' in the context of two Italian cities: Milan and Rome. Our findings suggest that perceived safety modulates the active population in an area, with effects that depend on age and gender. The overall effect of the appearance of safety in activity appears to be positive, even after controlling for population, employment density, and distance to the city center. Yet, the effect does not appear to be universal, and depends on the demographic with the population, with females and people older than 50 appearing to have a stronger preference for the appearance of safety.

Our results, however, do not provide a causal explanation of the observed effects. For instance, our data cannot distinguish between the hypotheses that people over than 50 prefer safer looking places, or that they modify their homes and shops to make the places they live and work at look safer. Nevertheless, they provide preliminary evidence suggesting a connection between the appearance of safety and levels of human activity that is strong enough to manifest itself at the city scale.

These methods, which could be readily applied to other cities if the data were available, can help improve modern efforts to use computational methods for urban recommendations. Some recent literature has focused on developing algorithms to recommend places for new business by using data on the presence of amenities in neighborhoods \cite{hidalgo2015we} and other urban features \cite{denadai2016death}.

\section{Acknowledgments}
Most of the computation of this article was done using free software and we are indebted to the developers and maintainers of the following packages: python, pandas, scikits.statsmodels, pysal to mention only a few.

\bibliographystyle{abbrv}
\bibliography{sigproc}  

\begin{thebibliography}{10}

\bibitem{arandjelovic2015netvlad}
R.~Arandjelovi{\'c}, P.~Gronat, A.~Torii, T.~Pajdla, and J.~Sivic.
\newblock Net{V}{L}{A}{D}: {C}{N}{N} architecture for weakly supervised place
  recognition.
\newblock {\em arXiv preprint arXiv:1511.07247}, 2015.

\bibitem{arietta2014city}
S.~M. Arietta, A.~A. Efros, R.~Ramamoorthi, and M.~Agrawala.
\newblock City forensics: {U}sing visual elements to predict non-visual city
  attributes.
\newblock {\em IEEE TVCG}, 20(12):2624--2633, 2014.

\bibitem{corman2005carrots}
H.~Corman and N.~Mocan.
\newblock Carrots, sticks, and broken windows.
\newblock {\em Journal of Law and Economics}, 48(1):235--266, 2005.

\bibitem{denadai2016death}
M.~De~Nadai, J.~Staiano, R.~Larcher, N.~Sebe, D.~Quercia, and B.~Lepri.
\newblock The {D}eath and {L}ife of {G}reat {I}talian {C}ities: {A} {M}obile
  {P}hone {D}ata {P}erspective.
\newblock In {\em ACM WWW}, 2016.

\bibitem{doeksen1997reducing}
H.~Doeksen.
\newblock Reducing crime and the fear of crime by reclaiming {N}ew {Z}ealand's
  suburban street.
\newblock {\em Landscape and urban planning}, 39(2):243--252, 1997.

\bibitem{doersch2015makes}
C.~Doersch, S.~Singh, A.~Gupta, J.~Sivic, and A.~A. Efros.
\newblock What makes {P}aris look like {P}aris?
\newblock {\em Communications of the ACM}, 58(12):103--110, 2015.

\bibitem{dubey16eccv}
A.~Dubey, N.~Naik, D.~Parikh, R.~Raskar, and C.~A. Hidalgo.
\newblock Deep learning the city: Quantifying urban perception at a global
  scale.
\newblock {\em ECCV}, 2016.

\bibitem{eagle2010}
N.~Eagle, M.~Macy, and R.~Claxton.
\newblock Network diversity and economic development.
\newblock {\em Science}, 328(5981):1029--1031, 2010.

\bibitem{felson1998opportunity}
M.~Felson and R.~V. Clarke.
\newblock Opportunity makes the thief.
\newblock {\em Police research series, paper}, 98, 1998.

\bibitem{glaeser2015big}
E.~L. Glaeser, S.~D. Kominers, M.~Luca, and N.~Naik.
\newblock Big data and big cities: The promises and limitations of improved
  measures of urban life.
\newblock Working Paper 21778, National Bureau of Economic Research, 2015.

\bibitem{gonzalez2008}
M.~Gonzalez, C.~Hidalgo, and L.~Barabasi.
\newblock Understanding individual mobility patterns.
\newblock {\em Nature}, 453(7196):779--782, 2008.

\bibitem{grabner2013visual}
H.~Grabner, F.~Nater, M.~Druey, and L.~Van~Gool.
\newblock Visual interestingness in image sequences.
\newblock In {\em ACM Multimedia}, 2013.

\bibitem{harcourt2009illusion}
B.~E. Harcourt.
\newblock {\em Illusion of order: The false promise of broken windows
  policing}.
\newblock Harvard University Press, 2009.

\bibitem{harcourt2006broken}
B.~E. Harcourt and J.~Ludwig.
\newblock Broken windows: New evidence from new york city and a five-city
  social experiment.
\newblock {\em The University of Chicago Law Review}, pages 271--320, 2006.

\bibitem{harcourt2007reefer}
B.~E. Harcourt and J.~Ludwig.
\newblock Reefer madness: Broken windows policing and misdemeanor marijuana
  arrests in new york city, 1989-2000.
\newblock {\em Criminology and Public Policy}, 2007.

\bibitem{herbrich2006trueskill}
R.~Herbrich, T.~Minka, and T.~Graepel.
\newblock Trueskill™: {A} {B}ayesian skill rating system.
\newblock In {\em NIPS}, 2006.

\bibitem{hidalgo2015we}
C.~A. Hidalgo and E.~E. Casta{\~n}er.
\newblock Do we need another coffee house? {T}he amenity space and the
  evolution of neighborhoods.
\newblock {\em arXiv preprint arXiv:1509.02868}, 2015.

\bibitem{hidalgo2008dynamics}
C.~A. Hidalgo and C.~Rodriguez-Sickert.
\newblock The dynamics of a mobile phone network.
\newblock {\em Physica A: Statistical Mechanics and its Applications},
  387(12):3017--3024, 2008.

\bibitem{hollway1997risk}
W.~Hollway and T.~Jefferson.
\newblock The risk society in an age of anxiety: situating fear of crime.
\newblock {\em British journal of sociology}, pages 255--266, 1997.

\bibitem{isaacman2010tale}
S.~Isaacman, R.~Becker, R.~C{\'a}ceres, S.~Kobourov, J.~Rowland, and
  A.~Varshavsky.
\newblock A tale of two cities.
\newblock In {\em ACM HotMobile}, pages 19--24, 2010.

\bibitem{isola2011understanding}
P.~Isola, D.~Parikh, A.~Torralba, and A.~Oliva.
\newblock Understanding the intrinsic memorability of images.
\newblock In {\em NIPS}, 2011.

\bibitem{jacobs1961death}
J.~Jacobs.
\newblock {\em The death and life of {A}merican cities}.
\newblock Random House, 1961.

\bibitem{jacobsdefensible}
J.~M. Jacobs and L.~Lees.
\newblock Defensible space on the move: revisiting the urban geography of alice
  coleman.
\newblock In {\em International Journal of Urban and Regional Research 37.5
  (2013): 1559-1583}, 2013.

\bibitem{jia2014caffe}
Y.~Jia, E.~Shelhamer, J.~Donahue, S.~Karayev, J.~Long, R.~Girshick,
  S.~Guadarrama, and T.~Darrell.
\newblock Caffe: {C}onvolutional {A}rchitecture for {F}ast {F}eature
  {E}mbedding.
\newblock {\em arXiv preprint arXiv:1408.5093}, 2014.

\bibitem{keizer2008spreading}
K.~Keizer, S.~Lindenberg, and L.~Steg.
\newblock The spreading of disorder.
\newblock {\em Science}, 322(5908):1681--1685, 2008.

\bibitem{kelling1997fixing}
G.~L. Kelling and C.~M. Coles.
\newblock {\em Fixing broken windows: {R}estoring order and reducing crime in
  our communities}.
\newblock Simon and Schuster, 1997.

\bibitem{kelling2001police}
G.~L. Kelling and W.~H. Sousa.
\newblock {\em Do Police Matter?: An Analysis of the Impact of New York City's
  Police Reforms}.
\newblock CCI Center for Civic Innovation at the Manhattan Institute, 2001.

\bibitem{khosla2014looking}
A.~Khosla, B.~An, J.~J. Lim, and A.~Torralba.
\newblock Looking beyond the visible scene.
\newblock In {\em CVPR}, 2014.

\bibitem{krizhevsky2012imagenet}
A.~Krizhevsky, I.~Sutskever, and G.~E. Hinton.
\newblock Imagenet classification with deep convolutional neural networks.
\newblock In {\em NIPS}, 2012.

\bibitem{Lenormand150449}
M.~Lenormand, M.~Picornell, O.~G. Cant{\'u}-Ros, T.~Louail, R.~Herranz,
  M.~Barthelemy, E.~Fr{\'\i}as-Mart{\'\i}nez, M.~San~Miguel, and J.~J. Ramasco.
\newblock Comparing and modelling land use organization in cities.
\newblock {\em Royal Society Open Science}, 2(12), 2015.

\bibitem{lynch1960image}
K.~Lynch.
\newblock {\em The image of the city}, volume~11.
\newblock MIT press, 1960.

\bibitem{mark1984fear}
W.~Mark.
\newblock Fear of victimization: {W}hy are women and the elderly more afraid?
\newblock {\em Social science quarterly}, 65(3):681, 1984.

\bibitem{milgram1970experience}
S.~Milgram.
\newblock The experience of living in cities.
\newblock {\em Science}, 167(3924):1461, 1970.

\bibitem{naik2015people}
N.~Naik, S.~D. Kominers, R.~Raskar, E.~L. Glaeser, and C.~A. Hidalgo.
\newblock Do people shape cities, or do cities shape people? {T}he co-evolution
  of physical, social, and economic change in five major {U}{S} cities.
\newblock Technical report, National Bureau of Economic Research, 2015.

\bibitem{naik2014streetscore}
N.~Naik, J.~Philipoom, R.~Raskar, and C.~Hidalgo.
\newblock Streetscore--predicting the perceived safety of one million
  streetscapes.
\newblock In {\em CVPRW}, 2014.

\bibitem{naik2016cities}
N.~Naik, R.~Raskar, and C.~A. Hidalgo.
\newblock Cities are physical too: Using computer vision to measure the quality
  and impact of urban appearance.
\newblock {\em The American Economic Review}, 106(5):128--132, 2016.

\bibitem{nasar1998evaluative}
J.~L. Nasar.
\newblock {\em The evaluative image of the city}.
\newblock Sage Publications Thousand Oaks, CA, 1998.

\bibitem{nasar1993proximate}
J.~L. Nasar, B.~Fisher, and M.~Grannis.
\newblock Proximate physical cues to fear of crime.
\newblock {\em Landscape and urban planning}, 26(1):161--178, 1993.

\bibitem{newman1972defensible}
O.~Newman.
\newblock {\em Defensible space}.
\newblock Macmillan New York, 1972.

\bibitem{oquab2014learning}
M.~Oquab, L.~Bottou, I.~Laptev, and J.~Sivic.
\newblock Learning and transferring mid-level image representations using
  convolutional neural networks.
\newblock In {\em CVPR}, 2014.

\bibitem{ordonez2014learning}
V.~Ordonez and T.~L. Berg.
\newblock Learning high-level judgments of urban perception.
\newblock In {\em ECCV}, 2014.

\bibitem{pantazis2000fear}
C.~Pantazis.
\newblock `{F}ear of {C}rime’, {V}ulnerability and {P}overty.
\newblock {\em British journal of criminology}, 40(3):414--436, 2000.

\bibitem{porzi2015predicting}
L.~Porzi, S.~Rota~Bul{\`o}, B.~Lepri, and E.~Ricci.
\newblock Predicting and {U}nderstanding {U}rban {P}erception with
  {C}onvolutional {N}eural {N}etworks.
\newblock In {\em ACM Multimedia}, 2015.

\bibitem{quercia2014aesthetic}
D.~Quercia, N.~K. O'Hare, and H.~Cramer.
\newblock Aesthetic capital: what makes {L}ondon look beautiful, quiet, and
  happy?
\newblock In {\em ACM CSCW}, 2014.

\bibitem{salesses2013collaborative}
P.~Salesses, K.~Schechtner, and C.~A. Hidalgo.
\newblock The collaborative image of the city: mapping the inequality of urban
  perception.
\newblock {\em PloS one}, 8(7):e68400, 2013.

\bibitem{sampson2001disorder}
R.~J. Sampson and S.~W. Raudenbush.
\newblock {\em Disorder in urban neighborhoods: Does it lead to crime}.
\newblock US Department of Justice, Office of Justice Programs, National
  Institute of Justice, 2001.

\bibitem{sampson2004seeing}
R.~J. Sampson and S.~W. Raudenbush.
\newblock Seeing disorder: Neighborhood stigma and the social construction of
  “broken windows”.
\newblock {\em Social psychology quarterly}, 67(4):319--342, 2004.

\bibitem{sampson1997neighborhoods}
R.~J. Sampson, S.~W. Raudenbush, and F.~Earls.
\newblock Neighborhoods and violent crime: A multilevel study of collective
  efficacy.
\newblock {\em Science}, 277(5328):918--924, 1997.

\bibitem{sartori2015affective}
A.~Sartori, V.~Yanulevskaya, A.~A. Salah, J.~Uijlings, E.~Bruni, and N.~Sebe.
\newblock Affective analysis of professional and amateur abstract paintings
  using statistical analysis and art theory.
\newblock {\em ACM TiiS}, 5(2):8, 2015.

\bibitem{simonyan2014very}
K.~Simonyan and A.~Zisserman.
\newblock Very deep convolutional networks for large-scale image recognition.
\newblock {\em arXiv preprint arXiv:1409.1556}, 2014.

\bibitem{sivic2014urban}
J.~Sivic and A.~A. Efros.
\newblock Urban-{S}cale {Q}uantitative {V}isual {A}nalysis.
\newblock {\em Smart Cities}, page~43, 2014.

\bibitem{taylor1986testing}
R.~B. Taylor and M.~Hale.
\newblock Testing alternative models of fear of crime.
\newblock {\em The Journal of Criminal Law and Criminology (1973-)},
  77(1):151--189, 1986.

\bibitem{tiefelsdorf2007semiparametric}
M.~Tiefelsdorf and D.~A. Griffith.
\newblock Semiparametric filtering of spatial autocorrelation: the eigenvector
  approach.
\newblock {\em Environment and Planning A}, 39(5):1193--1221, 2007.

\bibitem{wekerle1995safe}
G.~R. Wekerle and C.~Whitzman.
\newblock {\em Safe cities: {G}uidelines for planning, design, and management}.
\newblock Van Nostrand Reinhold Company, 1995.

\bibitem{wilson1982broken}
J.~Q. Wilson and G.~L. Kelling.
\newblock Broken windows.
\newblock {\em Critical issues in policing: Contemporary readings}, pages
  395--407, 1982.

\bibitem{zeiler2014visualizing}
M.~D. Zeiler and R.~Fergus.
\newblock Visualizing and understanding convolutional networks.
\newblock In {\em ECCV}, 2014.

\bibitem{zhou2014learning}
B.~Zhou, A.~Lapedriza, J.~Xiao, A.~Torralba, and A.~Oliva.
\newblock Learning deep features for scene recognition using places database.
\newblock In {\em NIPS}, 2014.

\end{thebibliography}
%
%
\balancecolumns

\end{document}